\documentclass[prb,twocolumn,superscriptaddress,showpacs,amsmath,amssymb,longbibliography]{revtex4-2}
\usepackage{soul}
\usepackage{pifont}
\usepackage{verbatim}
\usepackage{graphicx}%
\usepackage{dcolumn}%
\usepackage{bm}%
\usepackage{xcolor}
\usepackage{url}
\usepackage[normalem]{ulem} %
\usepackage{cancel}
\usepackage{relsize}

\definecolor{goodred}{rgb}{0.7,0,0}
\usepackage[colorlinks,urlcolor=blue,citecolor=blue,linkcolor=goodred]{hyperref}
\usepackage[capitalise]{cleveref}

\usepackage{physics}
\usepackage{amsthm,amsbsy,amsfonts,gensymb}
\usepackage{wrapfig}

\newcommand{\I}{\mathbb{I}}

\newcommand{\Ts}{\mathcal{T}}
\newcommand{\mTs}{$\Ts$}
\newcommand{\SITs}{\mathcal{K}}
\newcommand{\QGT}{\mathcal{Q}}

\usepackage[utf8]{inputenc}
\usepackage{pgfplots}
\usepgfplotslibrary{groupplots,dateplot}
\usetikzlibrary{patterns,shapes.arrows,math,calc,arrows.meta}
\pgfplotsset{compat=newest}
\usepackage{tikz-3dplot}

\usetikzlibrary{external}
\usetikzlibrary{positioning}

\iftrue
\usepackage{pdfpages}
\usepackage{pgffor}
\makeatletter
\AtBeginDocument{\let\LS@rot\@undefined}
\makeatother
\fi

\begin{document}
\title{Quantum Geometric Helical Superconductivity}

\author{Aaron Dunbrack}
\affiliation{Department of Physics and Nanoscience Center, University of Jyv\"askyl\"a, P.O. Box 35, FI-40014 University of Jyv\"askyl\"a, Finland}

\author{Pauli Virtanen}
\affiliation{Department of Physics and Nanoscience Center, University of Jyv\"askyl\"a, P.O. Box 35, FI-40014 University of Jyv\"askyl\"a, Finland}

\author{Tero T. Heikkil\"a}
\affiliation{Department of Physics and Nanoscience Center, University of Jyv\"askyl\"a, P.O. Box 35, FI-40014 University of Jyv\"askyl\"a, Finland}

\date{\today}

\begin{abstract}
Several physical phenomena in superconductors, such as helical superconductivity and the diode effect, rely on breaking time-reversal symmetry. This symmetry-breaking is usually accounted for via the Lifshitz invariant, a contribution to the free energy which is linear in the phase gradient of the order parameter. In dispersive single-band superconductors with conventional pairing, the Lifshitz invariant can be computed from the asymmetries of the spectrum near the Fermi surface. We show that in multi-band superconductors, the quantum geometry also contributes to the Lifshitz invariant, and this is the dominant contribution when the low-energy bands are flat. We also analogously demonstrate quantum-geometry-driven commensurate-incommensurate transitions in charge and pair density waves.
\end{abstract}

\maketitle

\section{Introduction}\label{Sec:Intro}

In the past few years there has been much focus on superconductivity in flat bands, where a Fermi surface description breaks down. This is motivated in part by the discovery of superconductivity in magic angle twisted bilayer graphene. Proper description of the physics in this limit requires accounting for features of the electron bands other than the dispersion close to the Fermi level.
Central in these new descriptions is a dependence on \textit{quantum geometry}: it describes the eigenvectors of the Hamiltonian rather than just the eigenvalues, and in particular on the variation of the eigenvectors with the crystal momentum $k$. Under suitable assumptions --- notably including time-reversal symmetry --- the superfluid weight in a flat band is directly proportional to the quantum metric, and it receives a contribution from the quantum metric even in a dispersive band \cite{peotta2015superfluidity,OrigQMSFWeight,MinQuantMetric,PhysRevB.98.220511,PhysRevB.101.060505,PhysRevB.102.184504,PhysRevB.106.184507,PhysRevA.107.023313,PhysRevB.107.224505,PhysRevB.109.214518,PhysRevB.109.174508,PhysRevB.110.094505,PhysRevLett.132.026002,PhysRevResearch.6.L022053,PhysRevResearch.6.013256,PhysRevB.104.L100501,PhysRevLett.126.027002,PhysRevLett.128.087002,PhysRevB.108.094508,PhysRevLett.131.016002,chen2023pair}. What happens to these contributions without time-reversal symmetry has heretofore remained largely unexplored.

On general symmetry grounds, superconductors without time-reversal symmetry allow for several new phenomena, particularly when combined with inversion symmetry breaking. In terms of concrete applications, considerable attention has been paid to \textit{diode effects} \cite{SDERef1} where the critical current (above which the superconductivity breaks down) is higher in one direction of current flow than the other. More theoretically, time-reversal symmetry breaking makes a \textit{helical superconductor} \cite{HelicalRef1,HelicalRef2,HelicalRef3,agterberg2012book}, where the superconducting order parameter takes on a phase modulation.

\begin{figure}
    \centering
    \includegraphics[]{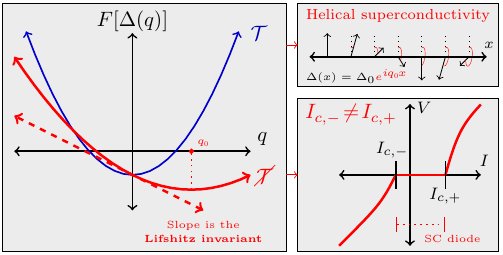}

    \caption{[Left] The phase gradient $q$ of the superconducting order parameter $\Delta$ is odd under time-reversal symmetry \mTs. Accordingly, for time-reversal symmetric systems (blue), the free energy $F$ is an even function of $q$, and thus is quadratic in phase gradients; this quadratic coefficient is the \textit{superfluid stiffness}. However, if time-reversal symmetry is broken (red), then the free energy can have a linear coefficient called the \textit{Lifshitz invariant}, which quantifies [top-right] helical superconductivity and [bottom-right] the superconducting diode effect.}
    
    \label{fig:TRSBreakingOverview}
\end{figure}

These effects can be quantified in terms of the \textit{Lifshitz invariant} \cite{agterberg2012book,mineev1994,edelstein1996}, the coefficient of the contribution to the free energy that is linear in phase gradient. In other words, if $\Psi$ is the superconducting order parameter, and one considers the part of free energy
\begin{equation}
    F[\Psi(x)]=\vec d\cdot (i\Psi^\dagger\vec\nabla\Psi+h.c.)+\ldots
\end{equation}
then $\vec d$ is the Lifshitz invariant, as illustrated in \cref{fig:TRSBreakingOverview}.

The quantum geometric contributions to the Lifshitz invariant have been considered to some limited extent. However, perhaps the most central question remains unanswered: what are the consequences of breaking time-reversal symmetry in the quantum geometry itself? Ref.~\onlinecite{QG_AnaSup} breaks time-reversal by adding a suitable pairing term, and Ref.~\onlinecite{QG_SDE} considers quantum geometric effects when the spectrum breaks time-reversal symmetry, but both still preserve the symmetry at the level of the normal-state wavefunctions.

In this paper, we resolve this problem for a wide range of Hamiltonians. We consider the case where time-reversal is broken only weakly, introduce a small parameter $\alpha$ to characterize this perturbation, and use Ginzburg-Landau (GL) theory to compute the Lifshitz invariant. We achieve this by extending the notion of quantum geometry in momentum space to include an extra dimension characterized by the parameter $\alpha$. We then apply this method to flat bands on certain bipartite models. Finally, we speculate on systems where this formalism could be applied, particularly within the flat-band systems of twisted and rhombohedral graphene \cite{OrigTBGSC,RTriGSC,RTetraGSC1,RTetraGSC2}, where time-reversal-symmetry-breaking effects have been observed \cite{RTriGSC,RTetraGSC1,RTetraGSC2,TBGSDE1,TBGSDE2,TTGSDE}, and briefly describe the extension of these methods to other strongly-interacting phases containing density waves and to nonstandard superconducting order parameters (discussed further in \cref{Apx:DWstates,Apx:OtherOrders}).

\section{Model requirements\label{Sec:ModelReqs}}
\begin{figure}
    \centering
    \includegraphics[]{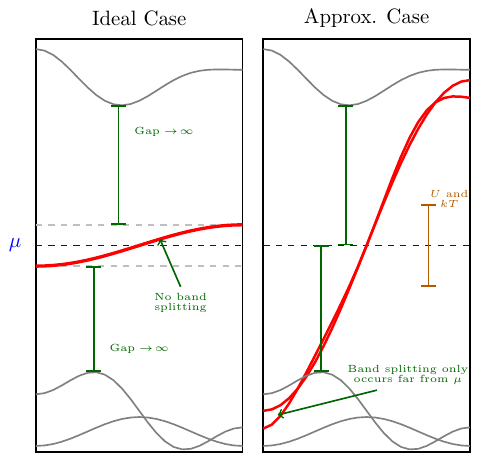}

    \caption{To characterize superconductivity in terms of abelian quantum geometry, the bandstructure needs to decompose into low-energy bands (red) near the Fermi level $\mu$ (blue), and high-energy bands which are far enough from it to be integrated out. The approximations are illustrated in green: in the simplest case (left), the low-energy bands are degenerate with each other and isolated from other bands by a large gap. These requirements can be relaxed somewhat (right); the methods presented here are a good approximation provided both the high-energy bands and any energy-splittings between the low-energy bands are located far from the Fermi energy compared to $U$ and $k_BT$.}
    
    \label{fig:bandreqs}
\end{figure}

In its simplest setting, the connection between superconductivity and quantum geometry involves decomposing the spectrum into low and high-energy bands, whereupon the projectors onto the low-energy bands yield quantum geometric quantities. To present a straightforward analytic calculation, we make several simplifying assumptions about the model, with or without time-reversal symmetry: (1) the spectrum separates into degenerate low-energy bands and widely-gapped high-energy bands (see \cref{fig:bandreqs}); (2) the paired electrons are time-reversed partners, implying spin singlet pairing; and (3) within the low-energy bands, pairing is of uniform strength, a criterion known as the \textit{uniform pairing condition} \cite{peotta2015superfluidity,OrigQMSFWeight,MinQuantMetric}.

The assumptions constrain the noninteracting part of the Hamiltonian under consideration, but the first condition can be relaxed. Provided (1a) the high-energy bands are always further from the Fermi surface than the interaction or thermal energies and (1b) the low-energy bands are degenerate when near the Fermi surface, the methods presented here are still a good approximation. The exact and approximate cases are compared in \cref{fig:bandreqs}. For simplicity, we consider here only an exactly flat band at the chemical potential, and relegate discussion of dispersive contributions --- including a mixed geometric-dispersive contribution to superfluid weight --- to \cref{Apx:QGDisp}.

We characterize the time-reversal symmetry breaking by a small parameter $\alpha$ which can be continuously tuned to a time-reversal symmetric Hamiltonian at $\alpha=0$. For the full perturbative expansion to be valid, the above assumptions on the bandstructure and pairing should hold for all values of $\alpha$, but this requirement can be neglected for the linear-order-in-$\alpha$ contribution to the Lifshitz invariant.

\section{Quantum Geometry in Superconductors\label{Sec:QGinSC}}
We now present the relation between superconductivity and the quantum metric in the presence of time-reversal symmetry. As derived in \cref{Apx:FreeEnResult}, in a flat band, the free energy expanded to quadratic order in the order parameter is given by
\begin{equation}\begin{split}\label{eq:FreeEnResult}
    \Omega^{(2)}=\sum_q|\Delta_0(q)|^2\Big\{\frac{N}{U}-\frac{\sum_{k}\Tr[P_k\Ts P_{-k-q}\Ts^{-1}]}{8T}\Bigg\}
\end{split}\end{equation}
where $\Delta_0$ is the order parameter, $U$ is the interaction strength, $N$ is the number of low-energy bands, $T$ is the temperature, $P$ is the projector onto the low-energy bands, and \mTs\ is the time-reversal operator.

The portion of this equation that relates to quantum geometry is the projector trace in the second term. Because of time-reversal symmetry, $\Ts P_{-k-q}\Ts^{-1}=P_{k+q}$, whereupon the trace evaluates to
\begin{equation}\label{eq:QMinprojTRS}
    \Tr[P_kP_{k+q}]=1-g_{ij}q^iq^j/2+O(q^3),
\end{equation}
where $g_{ij}$ is the quantum metric (see \cref{Apx:QGReview} for a general overview of quantum geometry). For the purposes of Ginzburg-Landau theory, the superfluid weight $D^{ij}$ results from expanding the free energy as
\begin{equation}
    \Omega^{(2)}=\sum_q|\Delta_0(q)|^2\left\{a_0+D_{ij}q^iq^j/2+\ldots\right\}
\end{equation}
and therefore in a flat band the superfluid weight is directly proportional to the integrated quantum metric, $D_{ij}=\frac{1}{8T}\sum_k g_{ij}(k)$.

\subsection{Breaking time-reversal symmetry\label{Sec:BreakTRS}} We now remove the time-reversal symmetry with a perturbation. Specifically, we assume the Hamiltonian of the system under consideration is continuously parameterized by a $\Ts$-breaking parameter $\alpha$ as
\begin{equation}
    H(\alpha)=H_0(\alpha^2)+\alpha H_1(\alpha^2)
\end{equation}
where $[\Ts,H_0]=0=\{\Ts,H_1\}$. This decomposition of the Hamiltonian into its \mTs-symmetric and antisymmetric parts, which can always be done but demands a suitable definition of $\alpha$, ensures $\Ts H(\alpha)\Ts^{-1}=H(-\alpha)$. We denote the value of $\alpha$ corresponding to a physical choice of Hamiltonian as $\bar\alpha$.

The resulting projectors now depend on $\alpha$, expressed in the notation $P_{k,\alpha}$. The reason for assuming this particular form of the Hamiltonian is to ensure the property $\Ts P_{k,\alpha}\Ts^{-1}=P_{-k,-\alpha}$, which follows naturally from $\Ts H(\alpha)\Ts^{-1}=H(-\alpha)$. Strictly speaking the latter condition is not required provided the former is met, but the decomposition is natural so little is lost by assuming this slightly stronger condition.

The derivation of \cref{eq:FreeEnResult} performed in \cref{Apx:FreeEnResult} does not rely on time-reversal; only the simplification of the trace to \cref{eq:QMinprojTRS} does. In this case, the trace instead evaluates to
\begin{equation}\label{eq:GenTrinFreeEnnoTRS}
    \Tr[P_{k,\bar\alpha}\Ts P_{-k-q,\bar\alpha}\Ts^{-1}]=\Tr[P_{k,\bar\alpha}P_{k+q,-\bar\alpha}].
\end{equation}

\begin{figure}
    \includegraphics[]{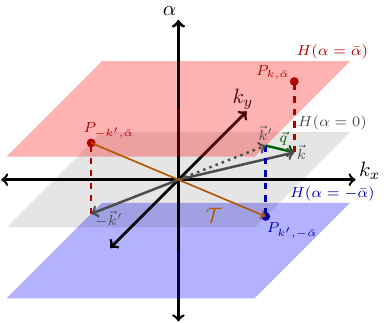}
    
    \caption{Instead of evaluating the quantum geometry in the usual $(k_x,k_y)$-space, we evaluate it in $(k_x,k_y,\alpha)$-space, where $\alpha$ is a perturbative \mTs-breaking parameter. The gray plane at $\alpha=0$ indicates the \mTs-symmetric case, the red plane indicates at finite \mTs-breaking $\alpha=\bar\alpha$, and the blue plane is the time-reverse of the red plane (by assumption on the definition of $\alpha$). For \mTs-symmetric SCs, $P_k$ and $\Ts P_{-k'}\Ts^{-1}$ differ only in $k$-space (by momentum vector $q$ from the gradient expansion), but without this symmetry, they also differ via $\alpha\rightarrow -\alpha$. Including this parameter in the quantum metric (and the quantum geometric coefficients of all higher-order corrections in $q$) then accounts for this difference.}
    
    \label{fig:QGinalpha}
\end{figure}

The two projectors can now be understood as separated both in momentum space and by their value of $\alpha$. In $d$ dimensions, this can be understood as being separated in a $d+1$-dimensional space, as illustrated in \cref{fig:QGinalpha}.

The earlier analysis can straightforwardly be extended to this case provided one takes a more general perspective on quantum geometry (discussed in \cref{Apx:QGReview}). For instance, to understand this expression to quadratic order in $q$ and $\bar\alpha$, one can use the extended parameter space illustrated in \cref{fig:QGinalpha} to express an extended quantum metric in terms of an index $\mu$ that runs both over spatial indices $i$ and an additional index $\alpha$,
\begin{equation}\label{eq:extendedQMdefn}
    g_{\mu\nu}=\begin{bmatrix}g_{ij}&g_{i\alpha}\\g_{\alpha j}&g_{\alpha\alpha}\end{bmatrix}=\Tr[\partial_\mu P\partial_\nu P],
\end{equation}
where $g_{ij}$ is the usual quantum metric in $k$-space. One could also introduce multiple \mTs-breaking parameters and view $\alpha$ as a vector in some multidimensional space without changing the underlying derivation.

With this extended quantum metric and the corresponding difference of parameters of the projectors in this extended space, $q^\mu=(q^i,2\bar\alpha)$, the expansion to combined quadratic order in $q$ and $\alpha$ (instead of \cref{eq:QMinprojTRS}) is
\begin{equation}\label{eq:QMinprojnoTRS}\begin{split}
    \Tr[P_kP_{k+q}]=&1-g_{\mu\nu}q^\mu q^\nu/2+\ldots\\
    =&(1-2\bar\alpha^2g_{\alpha\alpha})-2\bar\alpha g_{i\alpha}q^i-g_{ij}q^iq^j/2+\ldots
\end{split}\end{equation}

\section{Physical interpretation\label{Sec:PhysInterp}} The $g_{ij}$ term gives the usual geometric contribution to superfluid weight and $g_{\alpha\alpha}$ just describes a reduction in $T_c$ (see \cref{Apx:TC}), but the $g_{i\alpha}$ term is new and gives the Lifshitz invariant to linear order in $\alpha$. This off-diagonal component of the quantum metric quantifies the extent to which the derivatives of the projectors with respect to momentum and $\alpha$ couple to the same degrees of freedom. For example, if $\alpha$ is a uniform Zeeman field, a non-vanishing $g_{i\alpha}$ requires that momentum and spin are coupled as well, as proven in \cref{Apx:IndepPert}.

The value of $q$ which minimizes the free energy, known as the helical wavevector (illustrated in \cref{fig:TRSBreakingOverview}), is given by
\begin{equation}\label{eq:helicalwavevector}
    q_0=-2\bar\alpha\left[\int g_{ij}(k)dk\right]^{-1}\left[\int g_{i\alpha}(k)\right]
\end{equation}
where the terms in square brackets should be understood as a matrix and vector, respectively, in their $ij$ indices. In the flat band limit this purely quantum-geometric $q_0$ is independent of temperature, which differs from the dispersive case (see \cref{Apx:TempDep}).

One straightforward case to analyze is when $\Ts P_{-k}\Ts^{-1}=P_{k+2Q}$ -- i.e., when there is a effective pseudo-time-reversal $\Ts'$ that reflects momentum over a nonzero $Q$, corresponding to a pair density wave with perfect quantum geometric nesting \cite{QGNest}. In this case $\Tr[P_k\Ts P_{-k-q}\Ts^{-1}]$ takes its maximal possible value when $q=2Q$, which is an exact result of the perturbation theory (see \cref{Apx:PseudoTR}).

Generally speaking, higher-order corrections in $\alpha$ are similarly perfectly equivalent to higher-order corrections in $q$ --- the extension of $q^i$ to $q^\mu$ by the additional component $2\bar\alpha$ works to all orders of the gradient expansion. However, higher orders in the gradient expansion may require more complicated analogues of the minimal quantum metric condition (see \cref{Apx:MinQuantMet} and \cite{MinQuantMetric}), so this is a nontrivial extension.

\section{Bipartite model\label{Sec:Bipartite}}
\begin{figure}
    \centering
    \includegraphics[]{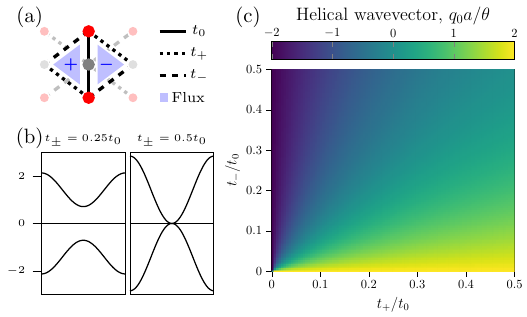}
    \caption{(a) 1d model with a bipartite flat band (one unit cell highlighted). Gray dots are A sites, red dots are B and C sites. Time-reversal symmetry is broken by the magnetic flux pattern in blue. (b) Sample spectra with no flux; the gap closes at $t_++t_-=t_0$. (c) Helical wavevector of the flat band, normalized by the flux-induced hopping phase $\theta$, as a function of hopping amplitudes $t_{\pm}$.}
    \label{fig:1dbipartite}
\end{figure}

We illustrate the time-reversal symmetry breaking quantum geometry with a tight-binding model on a lattice with three atoms (A,B,C) per unit cell and a bipartite structure such that all hoppings are from A atoms to either B or C atoms. Suppose the model features $I\Ts$ symmetry, where inversion symmetry $I$ interchanges the $B$ and $C$ sublattice and $\Ts=\SITs$. One example of such a lattice is shown in \cref{fig:1dbipartite}a, with spectra in \cref{fig:1dbipartite}b.

In the basis where $\psi=(\psi_A\ \psi_B\ \psi_C)^T$, the momentum-space Hamiltonian then takes the form
\begin{equation}\label{eq:genbipartiteham}
    H(k)=\begin{bmatrix}
        0&f(k)&f^*(k)\\
        f^*(k)&0&0\\
        f(k)&0&0
    \end{bmatrix}.
\end{equation}
The spectrum of this Hamiltonian features a flat band at $E=0$ with the wavefunction
\begin{equation}
    \psi_\text{FB}(k)=(0\ e^{-i\phi(k)/2}\ -\!e^{i\phi(k)/2})^T/\sqrt{2},
\end{equation}
where $\phi(k)=2\arg[f(k)]$. The dispersive bands have energies $\pm\sqrt{2}|f(k)|$, so to separate the flat band from the dispersive bands we assume $|f(k)|>0$ for all $k$. The projector onto the flat band is
\begin{equation}
    P_\text{FB}(k)=\frac{1}{2}\begin{bmatrix}
        0&0&0\\
        0&1&e^{i\phi(k)}\\
        0&e^{-i\phi(k)}&1
    \end{bmatrix}.
\end{equation}
Note the quantum geometry of the flat band would be unchanged by augmenting the model with $AA$ hoppings or symmetry-respecting on-site potentials.

At $\alpha=0$, time-reversal symmetry (or equivalently inversion, given $I\Ts$ symmetry) also implies $f(k)=f^*(-k)$. Therefore, to define our perturbative parameter $\alpha$, we decompose this phase as $\phi(\alpha,k)=\phi_0(k)+\alpha\tilde\phi(k)$, where $\phi_0(k)=-\phi_0(-k)$ and $\tilde\phi(k)=\tilde\phi(-k)$.

Taking the 1d model illustrated in \cref{fig:1dbipartite}a as a specific example, we evaluate the resulting $q_0$ in \cref{fig:1dbipartite}c (details in \cref{Apx:BipartiteEx}). When $t_+=t_-$ the system has a mirror symmetry $x\rightarrow -x$, and the Lifshitz invariant vanishes. When $t_{\pm}\rightarrow 0$, $q_0a$ is twice the phase of the hoppings, reflecting the existence of a gauge where all hoppings are real. In the limit where both $t_\pm\rightarrow 0$, the quantum geometry is trivial and the superfluid weight vanishes, giving nonanalytic behavior.

\section{Discussion\label{Sec:Discussion}}
We have demonstrated how to extend the derivation of the Ginzburg-Landau theory of superconductors in the presence of broken time-reversal symmetry at the level of the momentum dependent Bloch state wavefunctions. We do this as a perturbative expansion in the time-reversal symmetry breaking parameter $\alpha$.

To lowest order, the breaking of time-reversal and inversion symmetry is quantified by the Lifshitz invariant. It can be computed to leading order in the perturbative symmetry-breaking parameter from the off-diagonal components of the extended quantum metric, defined as the quantum metric in the combined $(k_x,k_y,\alpha)$ space. This then gives the helical wavevector $q_0$, which we computed for a class of bipartite models and plotted for an example model in \cref{fig:1dbipartite}.

We now speculate on the application of this approach to flat-band superconducting systems of interest, such as twisted and rhombohedral multilayer graphene. In general, these systems in their ideal form preserve $C_{3z}$ symmetry, therefore implying a vanishing Lifshitz invariant. However, a perturbation of the lattice such as strain can break this symmetry. Moreover, higher-odd-order terms are also directly accessible with this method, provided one describes higher-even-order terms first. This includes the $C_{3z}$-preserving $q^3$ term, as discussed for systems with time-reversal symmetric wavefunctions in Ref.~\onlinecite{QG_SDE}, computed from a generalization of the $q^4$ contribution for the time-reversal-symmetric case.

The most obvious candidates to show quantum geometric Lifshitz invariants are single-valley-polarized systems, such as the purported quarter-metal phase in rhombohedral tetralayer graphene \cite{RTetraGSC1} and similar phases in TBG \cite{SDERef1,PhysRevLett.130.266003}. Complete valley polarization strongly breaks time-reversal symmetry, but there should be relevant insights to be drawn from considering weak valley polarization via the perturbation theory presented here.

Moreover, both twisted and rhombohedral graphene systems also support superconducting phases which are likely to be time-reversal symmetric \cite{RTriGSC,RTetraGSC1,TBGSDE2}. If so, then a perturbation which breaks time-reversal symmetry could generate a quantum geometric Lifshitz invariant that is significant compared to the dispersive one. As a concrete example, a layer-dependent Zeeman field can couple to momentum in a way that provides a nonzero Lifshitz invariant even without spin-orbit coupling. Such a field can be generated by placing opposite-orientation ferromagnets on opposite sides of a graphene multilayer.

Finally, we note that mixed quantum geometry can also be used in a similar fashion beyond conventional superconductivity to compute effects of the wavefunction from parametric perturbations of the Hamiltonian. We show that one can apply the same formalism developed here to characterize a commensurate-incommensurate transition in charge and pair density wave states as they deviate from quantum geometric nesting \cite{QGNest} in \cref{Apx:DWstates}, and to changes in the form of the superconducting order parameter after breaking time-reversal symmetry in the presence of extended interactions (leading to unconventional superconductivity with intersite pairing) in \cref{Apx:OtherOrders}.

\nocite{CodeRef}

\section{Acknowledgments}
This work was supported by the Keele and Jane and Aatos Erkko foundations as part of the SuperC collaboration, by the Finnish Quantum Flagship (project no. 210000621611), and by grant NSF PHY-2309135 to the Kavli Institute for Theoretical Physics (KITP). This work is part of the Finnish Centre of Excellence in Quantum Materials (QMAT). The authors acknowledge useful conversations with Dan Crawford, Daniel Mu\~noz-Segovia, Dathan Ault-McCoy, P\"aivi T\"orm\"a and Stefan Ilic.

\bibliography{main_prb}

\begin{thebibliography}{44}%
\makeatletter
\providecommand \@ifxundefined [1]{%
 \@ifx{#1\undefined}
}%
\providecommand \@ifnum [1]{%
 \ifnum #1\expandafter \@firstoftwo
 \else \expandafter \@secondoftwo
 \fi
}%
\providecommand \@ifx [1]{%
 \ifx #1\expandafter \@firstoftwo
 \else \expandafter \@secondoftwo
 \fi
}%
\providecommand \natexlab [1]{#1}%
\providecommand \enquote  [1]{``#1''}%
\providecommand \bibnamefont  [1]{#1}%
\providecommand \bibfnamefont [1]{#1}%
\providecommand \citenamefont [1]{#1}%
\providecommand \href@noop [0]{\@secondoftwo}%
\providecommand \href [0]{\begingroup \@sanitize@url \@href}%
\providecommand \@href[1]{\@@startlink{#1}\@@href}%
\providecommand \@@href[1]{\endgroup#1\@@endlink}%
\providecommand \@sanitize@url [0]{\catcode `\\12\catcode `\$12\catcode `\&12\catcode `\#12\catcode `\^12\catcode `\_12\catcode `\%12\relax}%
\providecommand \@@startlink[1]{}%
\providecommand \@@endlink[0]{}%
\providecommand \url  [0]{\begingroup\@sanitize@url \@url }%
\providecommand \@url [1]{\endgroup\@href {#1}{\urlprefix }}%
\providecommand \urlprefix  [0]{URL }%
\providecommand \Eprint [0]{\href }%
\providecommand \doibase [0]{https://doi.org/}%
\providecommand \selectlanguage [0]{\@gobble}%
\providecommand \bibinfo  [0]{\@secondoftwo}%
\providecommand \bibfield  [0]{\@secondoftwo}%
\providecommand \translation [1]{[#1]}%
\providecommand \BibitemOpen [0]{}%
\providecommand \bibitemStop [0]{}%
\providecommand \bibitemNoStop [0]{.\EOS\space}%
\providecommand \EOS [0]{\spacefactor3000\relax}%
\providecommand \BibitemShut  [1]{\csname bibitem#1\endcsname}%
\let\auto@bib@innerbib\@empty
\bibitem [{\citenamefont {Peotta}\ and\ \citenamefont {T{\"o}rm{\"a}}(2015)}]{peotta2015superfluidity}%
  \BibitemOpen
  \bibfield  {author} {\bibinfo {author} {\bibfnamefont {S.}~\bibnamefont {Peotta}}\ and\ \bibinfo {author} {\bibfnamefont {P.}~\bibnamefont {T{\"o}rm{\"a}}},\ }\bibfield  {title} {\bibinfo {title} {Superfluidity in topologically nontrivial flat bands},\ }\href {https://doi.org/10.1038/ncomms9944} {\bibfield  {journal} {\bibinfo  {journal} {Nature communications}\ }\textbf {\bibinfo {volume} {6}},\ \bibinfo {pages} {8944} (\bibinfo {year} {2015})}\BibitemShut {NoStop}%
\bibitem [{\citenamefont {Liang}\ \emph {et~al.}(2017)\citenamefont {Liang}, \citenamefont {Vanhala}, \citenamefont {Peotta}, \citenamefont {Siro}, \citenamefont {Harju},\ and\ \citenamefont {T\"orm\"a}}]{OrigQMSFWeight}%
  \BibitemOpen
  \bibfield  {author} {\bibinfo {author} {\bibfnamefont {L.}~\bibnamefont {Liang}}, \bibinfo {author} {\bibfnamefont {T.~I.}\ \bibnamefont {Vanhala}}, \bibinfo {author} {\bibfnamefont {S.}~\bibnamefont {Peotta}}, \bibinfo {author} {\bibfnamefont {T.}~\bibnamefont {Siro}}, \bibinfo {author} {\bibfnamefont {A.}~\bibnamefont {Harju}},\ and\ \bibinfo {author} {\bibfnamefont {P.}~\bibnamefont {T\"orm\"a}},\ }\bibfield  {title} {\bibinfo {title} {Band geometry, {B}erry curvature, and superfluid weight},\ }\href {https://doi.org/10.1103/PhysRevB.95.024515} {\bibfield  {journal} {\bibinfo  {journal} {Phys. Rev. B}\ }\textbf {\bibinfo {volume} {95}},\ \bibinfo {pages} {024515} (\bibinfo {year} {2017})}\BibitemShut {NoStop}%
\bibitem [{\citenamefont {Huhtinen}\ \emph {et~al.}(2022)\citenamefont {Huhtinen}, \citenamefont {Herzog-Arbeitman}, \citenamefont {Chew}, \citenamefont {Bernevig},\ and\ \citenamefont {T\"orm\"a}}]{MinQuantMetric}%
  \BibitemOpen
  \bibfield  {author} {\bibinfo {author} {\bibfnamefont {K.-E.}\ \bibnamefont {Huhtinen}}, \bibinfo {author} {\bibfnamefont {J.}~\bibnamefont {Herzog-Arbeitman}}, \bibinfo {author} {\bibfnamefont {A.}~\bibnamefont {Chew}}, \bibinfo {author} {\bibfnamefont {B.~A.}\ \bibnamefont {Bernevig}},\ and\ \bibinfo {author} {\bibfnamefont {P.}~\bibnamefont {T\"orm\"a}},\ }\bibfield  {title} {\bibinfo {title} {Revisiting flat band superconductivity: Dependence on minimal quantum metric and band touchings},\ }\href {https://doi.org/10.1103/PhysRevB.106.014518} {\bibfield  {journal} {\bibinfo  {journal} {Phys. Rev. B}\ }\textbf {\bibinfo {volume} {106}},\ \bibinfo {pages} {014518} (\bibinfo {year} {2022})}\BibitemShut {NoStop}%
\bibitem [{\citenamefont {T\"orm\"a}\ \emph {et~al.}(2018)\citenamefont {T\"orm\"a}, \citenamefont {Liang},\ and\ \citenamefont {Peotta}}]{PhysRevB.98.220511}%
  \BibitemOpen
  \bibfield  {author} {\bibinfo {author} {\bibfnamefont {P.}~\bibnamefont {T\"orm\"a}}, \bibinfo {author} {\bibfnamefont {L.}~\bibnamefont {Liang}},\ and\ \bibinfo {author} {\bibfnamefont {S.}~\bibnamefont {Peotta}},\ }\bibfield  {title} {\bibinfo {title} {Quantum metric and effective mass of a two-body bound state in a flat band},\ }\href {https://doi.org/10.1103/PhysRevB.98.220511} {\bibfield  {journal} {\bibinfo  {journal} {Phys. Rev. B}\ }\textbf {\bibinfo {volume} {98}},\ \bibinfo {pages} {220511} (\bibinfo {year} {2018})}\BibitemShut {NoStop}%
\bibitem [{\citenamefont {Julku}\ \emph {et~al.}(2020)\citenamefont {Julku}, \citenamefont {Peltonen}, \citenamefont {Liang}, \citenamefont {Heikkil\"a},\ and\ \citenamefont {T\"orm\"a}}]{PhysRevB.101.060505}%
  \BibitemOpen
  \bibfield  {author} {\bibinfo {author} {\bibfnamefont {A.}~\bibnamefont {Julku}}, \bibinfo {author} {\bibfnamefont {T.~J.}\ \bibnamefont {Peltonen}}, \bibinfo {author} {\bibfnamefont {L.}~\bibnamefont {Liang}}, \bibinfo {author} {\bibfnamefont {T.~T.}\ \bibnamefont {Heikkil\"a}},\ and\ \bibinfo {author} {\bibfnamefont {P.}~\bibnamefont {T\"orm\"a}},\ }\bibfield  {title} {\bibinfo {title} {Superfluid weight and {B}erezinskii-{K}osterlitz-{T}houless transition temperature of twisted bilayer graphene},\ }\href {https://doi.org/10.1103/PhysRevB.101.060505} {\bibfield  {journal} {\bibinfo  {journal} {Phys. Rev. B}\ }\textbf {\bibinfo {volume} {101}},\ \bibinfo {pages} {060505} (\bibinfo {year} {2020})}\BibitemShut {NoStop}%
\bibitem [{\citenamefont {Wang}\ \emph {et~al.}(2020)\citenamefont {Wang}, \citenamefont {Chaudhary}, \citenamefont {Chen},\ and\ \citenamefont {Levin}}]{PhysRevB.102.184504}%
  \BibitemOpen
  \bibfield  {author} {\bibinfo {author} {\bibfnamefont {Z.}~\bibnamefont {Wang}}, \bibinfo {author} {\bibfnamefont {G.}~\bibnamefont {Chaudhary}}, \bibinfo {author} {\bibfnamefont {Q.}~\bibnamefont {Chen}},\ and\ \bibinfo {author} {\bibfnamefont {K.}~\bibnamefont {Levin}},\ }\bibfield  {title} {\bibinfo {title} {Quantum geometric contributions to the {BKT} transition: Beyond mean field theory},\ }\href {https://doi.org/10.1103/PhysRevB.102.184504} {\bibfield  {journal} {\bibinfo  {journal} {Phys. Rev. B}\ }\textbf {\bibinfo {volume} {102}},\ \bibinfo {pages} {184504} (\bibinfo {year} {2020})}\BibitemShut {NoStop}%
\bibitem [{\citenamefont {Kitamura}\ \emph {et~al.}(2022)\citenamefont {Kitamura}, \citenamefont {Daido},\ and\ \citenamefont {Yanase}}]{PhysRevB.106.184507}%
  \BibitemOpen
  \bibfield  {author} {\bibinfo {author} {\bibfnamefont {T.}~\bibnamefont {Kitamura}}, \bibinfo {author} {\bibfnamefont {A.}~\bibnamefont {Daido}},\ and\ \bibinfo {author} {\bibfnamefont {Y.}~\bibnamefont {Yanase}},\ }\bibfield  {title} {\bibinfo {title} {Quantum geometric effect on {F}ulde-{F}errell-{L}arkin-{O}vchinnikov superconductivity},\ }\href {https://doi.org/10.1103/PhysRevB.106.184507} {\bibfield  {journal} {\bibinfo  {journal} {Phys. Rev. B}\ }\textbf {\bibinfo {volume} {106}},\ \bibinfo {pages} {184507} (\bibinfo {year} {2022})}\BibitemShut {NoStop}%
\bibitem [{\citenamefont {Iskin}(2023{\natexlab{a}})}]{PhysRevA.107.023313}%
  \BibitemOpen
  \bibfield  {author} {\bibinfo {author} {\bibfnamefont {M.}~\bibnamefont {Iskin}},\ }\bibfield  {title} {\bibinfo {title} {Quantum-geometric contribution to the {B}ogoliubov modes in a two-band {B}ose-{E}instein condensate},\ }\href {https://doi.org/10.1103/PhysRevA.107.023313} {\bibfield  {journal} {\bibinfo  {journal} {Phys. Rev. A}\ }\textbf {\bibinfo {volume} {107}},\ \bibinfo {pages} {023313} (\bibinfo {year} {2023}{\natexlab{a}})}\BibitemShut {NoStop}%
\bibitem [{\citenamefont {Iskin}(2023{\natexlab{b}})}]{PhysRevB.107.224505}%
  \BibitemOpen
  \bibfield  {author} {\bibinfo {author} {\bibfnamefont {M.}~\bibnamefont {Iskin}},\ }\bibfield  {title} {\bibinfo {title} {Extracting quantum-geometric effects from {G}inzburg-{L}andau theory in a multiband {H}ubbard model},\ }\href {https://doi.org/10.1103/PhysRevB.107.224505} {\bibfield  {journal} {\bibinfo  {journal} {Phys. Rev. B}\ }\textbf {\bibinfo {volume} {107}},\ \bibinfo {pages} {224505} (\bibinfo {year} {2023}{\natexlab{b}})}\BibitemShut {NoStop}%
\bibitem [{\citenamefont {Jiang}\ and\ \citenamefont {Barlas}(2024)}]{PhysRevB.109.214518}%
  \BibitemOpen
  \bibfield  {author} {\bibinfo {author} {\bibfnamefont {G.}~\bibnamefont {Jiang}}\ and\ \bibinfo {author} {\bibfnamefont {Y.}~\bibnamefont {Barlas}},\ }\bibfield  {title} {\bibinfo {title} {Geometric superfluid weight of composite bands in multiorbital superconductors},\ }\href {https://doi.org/10.1103/PhysRevB.109.214518} {\bibfield  {journal} {\bibinfo  {journal} {Phys. Rev. B}\ }\textbf {\bibinfo {volume} {109}},\ \bibinfo {pages} {214518} (\bibinfo {year} {2024})}\BibitemShut {NoStop}%
\bibitem [{\citenamefont {Iskin}(2024)}]{PhysRevB.109.174508}%
  \BibitemOpen
  \bibfield  {author} {\bibinfo {author} {\bibfnamefont {M.}~\bibnamefont {Iskin}},\ }\bibfield  {title} {\bibinfo {title} {{C}ooper pairing, flat-band superconductivity, and quantum geometry in the pyrochlore-{H}ubbard model},\ }\href {https://doi.org/10.1103/PhysRevB.109.174508} {\bibfield  {journal} {\bibinfo  {journal} {Phys. Rev. B}\ }\textbf {\bibinfo {volume} {109}},\ \bibinfo {pages} {174508} (\bibinfo {year} {2024})}\BibitemShut {NoStop}%
\bibitem [{\citenamefont {Daido}\ \emph {et~al.}(2024)\citenamefont {Daido}, \citenamefont {Kitamura},\ and\ \citenamefont {Yanase}}]{PhysRevB.110.094505}%
  \BibitemOpen
  \bibfield  {author} {\bibinfo {author} {\bibfnamefont {A.}~\bibnamefont {Daido}}, \bibinfo {author} {\bibfnamefont {T.}~\bibnamefont {Kitamura}},\ and\ \bibinfo {author} {\bibfnamefont {Y.}~\bibnamefont {Yanase}},\ }\bibfield  {title} {\bibinfo {title} {Quantum geometry encoded to pair potentials},\ }\href {https://doi.org/10.1103/PhysRevB.110.094505} {\bibfield  {journal} {\bibinfo  {journal} {Phys. Rev. B}\ }\textbf {\bibinfo {volume} {110}},\ \bibinfo {pages} {094505} (\bibinfo {year} {2024})}\BibitemShut {NoStop}%
\bibitem [{\citenamefont {Chen}\ and\ \citenamefont {Law}(2024)}]{PhysRevLett.132.026002}%
  \BibitemOpen
  \bibfield  {author} {\bibinfo {author} {\bibfnamefont {S.~A.}\ \bibnamefont {Chen}}\ and\ \bibinfo {author} {\bibfnamefont {K.~T.}\ \bibnamefont {Law}},\ }\bibfield  {title} {\bibinfo {title} {{G}inzburg-{L}andau theory of flat-band superconductors with quantum metric},\ }\href {https://doi.org/10.1103/PhysRevLett.132.026002} {\bibfield  {journal} {\bibinfo  {journal} {Phys. Rev. Lett.}\ }\textbf {\bibinfo {volume} {132}},\ \bibinfo {pages} {026002} (\bibinfo {year} {2024})}\BibitemShut {NoStop}%
\bibitem [{\citenamefont {Wang}\ \emph {et~al.}(2024)\citenamefont {Wang}, \citenamefont {Assili},\ and\ \citenamefont {Kotetes}}]{PhysRevResearch.6.L022053}%
  \BibitemOpen
  \bibfield  {author} {\bibinfo {author} {\bibfnamefont {J.-A.}\ \bibnamefont {Wang}}, \bibinfo {author} {\bibfnamefont {M.}~\bibnamefont {Assili}},\ and\ \bibinfo {author} {\bibfnamefont {P.}~\bibnamefont {Kotetes}},\ }\bibfield  {title} {\bibinfo {title} {Topological superfluid responses of superconducting {D}irac semimetals},\ }\href {https://doi.org/10.1103/PhysRevResearch.6.L022053} {\bibfield  {journal} {\bibinfo  {journal} {Phys. Rev. Res.}\ }\textbf {\bibinfo {volume} {6}},\ \bibinfo {pages} {L022053} (\bibinfo {year} {2024})}\BibitemShut {NoStop}%
\bibitem [{\citenamefont {Tam}\ and\ \citenamefont {Peotta}(2024)}]{PhysRevResearch.6.013256}%
  \BibitemOpen
  \bibfield  {author} {\bibinfo {author} {\bibfnamefont {M.}~\bibnamefont {Tam}}\ and\ \bibinfo {author} {\bibfnamefont {S.}~\bibnamefont {Peotta}},\ }\bibfield  {title} {\bibinfo {title} {Geometry-independent superfluid weight in multiorbital lattices from the generalized random phase approximation},\ }\href {https://doi.org/10.1103/PhysRevResearch.6.013256} {\bibfield  {journal} {\bibinfo  {journal} {Phys. Rev. Res.}\ }\textbf {\bibinfo {volume} {6}},\ \bibinfo {pages} {013256} (\bibinfo {year} {2024})}\BibitemShut {NoStop}%
\bibitem [{\citenamefont {Ahn}\ and\ \citenamefont {Nagaosa}(2021)}]{PhysRevB.104.L100501}%
  \BibitemOpen
  \bibfield  {author} {\bibinfo {author} {\bibfnamefont {J.}~\bibnamefont {Ahn}}\ and\ \bibinfo {author} {\bibfnamefont {N.}~\bibnamefont {Nagaosa}},\ }\bibfield  {title} {\bibinfo {title} {Superconductivity-induced spectral weight transfer due to quantum geometry},\ }\href {https://doi.org/10.1103/PhysRevB.104.L100501} {\bibfield  {journal} {\bibinfo  {journal} {Phys. Rev. B}\ }\textbf {\bibinfo {volume} {104}},\ \bibinfo {pages} {L100501} (\bibinfo {year} {2021})}\BibitemShut {NoStop}%
\bibitem [{\citenamefont {Peri}\ \emph {et~al.}(2021)\citenamefont {Peri}, \citenamefont {Song}, \citenamefont {Bernevig},\ and\ \citenamefont {Huber}}]{PhysRevLett.126.027002}%
  \BibitemOpen
  \bibfield  {author} {\bibinfo {author} {\bibfnamefont {V.}~\bibnamefont {Peri}}, \bibinfo {author} {\bibfnamefont {Z.-D.}\ \bibnamefont {Song}}, \bibinfo {author} {\bibfnamefont {B.~A.}\ \bibnamefont {Bernevig}},\ and\ \bibinfo {author} {\bibfnamefont {S.~D.}\ \bibnamefont {Huber}},\ }\bibfield  {title} {\bibinfo {title} {Fragile topology and flat-band superconductivity in the strong-coupling regime},\ }\href {https://doi.org/10.1103/PhysRevLett.126.027002} {\bibfield  {journal} {\bibinfo  {journal} {Phys. Rev. Lett.}\ }\textbf {\bibinfo {volume} {126}},\ \bibinfo {pages} {027002} (\bibinfo {year} {2021})}\BibitemShut {NoStop}%
\bibitem [{\citenamefont {Herzog-Arbeitman}\ \emph {et~al.}(2022)\citenamefont {Herzog-Arbeitman}, \citenamefont {Peri}, \citenamefont {Schindler}, \citenamefont {Huber},\ and\ \citenamefont {Bernevig}}]{PhysRevLett.128.087002}%
  \BibitemOpen
  \bibfield  {author} {\bibinfo {author} {\bibfnamefont {J.}~\bibnamefont {Herzog-Arbeitman}}, \bibinfo {author} {\bibfnamefont {V.}~\bibnamefont {Peri}}, \bibinfo {author} {\bibfnamefont {F.}~\bibnamefont {Schindler}}, \bibinfo {author} {\bibfnamefont {S.~D.}\ \bibnamefont {Huber}},\ and\ \bibinfo {author} {\bibfnamefont {B.~A.}\ \bibnamefont {Bernevig}},\ }\bibfield  {title} {\bibinfo {title} {Superfluid weight bounds from symmetry and quantum geometry in flat bands},\ }\href {https://doi.org/10.1103/PhysRevLett.128.087002} {\bibfield  {journal} {\bibinfo  {journal} {Phys. Rev. Lett.}\ }\textbf {\bibinfo {volume} {128}},\ \bibinfo {pages} {087002} (\bibinfo {year} {2022})}\BibitemShut {NoStop}%
\bibitem [{\citenamefont {Porlles}\ and\ \citenamefont {Chen}(2023)}]{PhysRevB.108.094508}%
  \BibitemOpen
  \bibfield  {author} {\bibinfo {author} {\bibfnamefont {D.}~\bibnamefont {Porlles}}\ and\ \bibinfo {author} {\bibfnamefont {W.}~\bibnamefont {Chen}},\ }\bibfield  {title} {\bibinfo {title} {Quantum geometry of singlet superconductors},\ }\href {https://doi.org/10.1103/PhysRevB.108.094508} {\bibfield  {journal} {\bibinfo  {journal} {Phys. Rev. B}\ }\textbf {\bibinfo {volume} {108}},\ \bibinfo {pages} {094508} (\bibinfo {year} {2023})}\BibitemShut {NoStop}%
\bibitem [{\citenamefont {Jiang}\ and\ \citenamefont {Barlas}(2023)}]{PhysRevLett.131.016002}%
  \BibitemOpen
  \bibfield  {author} {\bibinfo {author} {\bibfnamefont {G.}~\bibnamefont {Jiang}}\ and\ \bibinfo {author} {\bibfnamefont {Y.}~\bibnamefont {Barlas}},\ }\bibfield  {title} {\bibinfo {title} {Pair density waves from local band geometry},\ }\href {https://doi.org/10.1103/PhysRevLett.131.016002} {\bibfield  {journal} {\bibinfo  {journal} {Phys. Rev. Lett.}\ }\textbf {\bibinfo {volume} {131}},\ \bibinfo {pages} {016002} (\bibinfo {year} {2023})}\BibitemShut {NoStop}%
\bibitem [{\citenamefont {Chen}\ and\ \citenamefont {Huang}(2023)}]{chen2023pair}%
  \BibitemOpen
  \bibfield  {author} {\bibinfo {author} {\bibfnamefont {W.}~\bibnamefont {Chen}}\ and\ \bibinfo {author} {\bibfnamefont {W.}~\bibnamefont {Huang}},\ }\bibfield  {title} {\bibinfo {title} {Pair density wave facilitated by {B}loch quantum geometry in nearly flat band multiorbital superconductors},\ }\href {https://doi.org/10.1007/s11433-023-2122-4} {\bibfield  {journal} {\bibinfo  {journal} {Science China Physics, Mechanics \& Astronomy}\ }\textbf {\bibinfo {volume} {66}},\ \bibinfo {pages} {287212} (\bibinfo {year} {2023})}\BibitemShut {NoStop}%
\bibitem [{\citenamefont {Nadeem}\ \emph {et~al.}(2023)\citenamefont {Nadeem}, \citenamefont {Fuhrer},\ and\ \citenamefont {Wang}}]{SDERef1}%
  \BibitemOpen
  \bibfield  {author} {\bibinfo {author} {\bibfnamefont {M.}~\bibnamefont {Nadeem}}, \bibinfo {author} {\bibfnamefont {M.~S.}\ \bibnamefont {Fuhrer}},\ and\ \bibinfo {author} {\bibfnamefont {X.}~\bibnamefont {Wang}},\ }\bibfield  {title} {\bibinfo {title} {The superconducting diode effect},\ }\href@noop {} {\bibfield  {journal} {\bibinfo  {journal} {Nature Reviews Physics}\ }\textbf {\bibinfo {volume} {5}},\ \bibinfo {pages} {558} (\bibinfo {year} {2023})}\BibitemShut {NoStop}%
\bibitem [{\citenamefont {Agterberg}\ and\ \citenamefont {Kaur}(2007)}]{HelicalRef1}%
  \BibitemOpen
  \bibfield  {author} {\bibinfo {author} {\bibfnamefont {D.~F.}\ \bibnamefont {Agterberg}}\ and\ \bibinfo {author} {\bibfnamefont {R.~P.}\ \bibnamefont {Kaur}},\ }\bibfield  {title} {\bibinfo {title} {Magnetic-field-induced helical and stripe phases in {R}ashba superconductors},\ }\href {https://doi.org/10.1103/PhysRevB.75.064511} {\bibfield  {journal} {\bibinfo  {journal} {Phys. Rev. B}\ }\textbf {\bibinfo {volume} {75}},\ \bibinfo {pages} {064511} (\bibinfo {year} {2007})}\BibitemShut {NoStop}%
\bibitem [{\citenamefont {Dimitrova}\ and\ \citenamefont {Feigel'man}(2007)}]{HelicalRef2}%
  \BibitemOpen
  \bibfield  {author} {\bibinfo {author} {\bibfnamefont {O.}~\bibnamefont {Dimitrova}}\ and\ \bibinfo {author} {\bibfnamefont {M.~V.}\ \bibnamefont {Feigel'man}},\ }\bibfield  {title} {\bibinfo {title} {Theory of a two-dimensional superconductor with broken inversion symmetry},\ }\href {https://doi.org/10.1103/PhysRevB.76.014522} {\bibfield  {journal} {\bibinfo  {journal} {Phys. Rev. B}\ }\textbf {\bibinfo {volume} {76}},\ \bibinfo {pages} {014522} (\bibinfo {year} {2007})}\BibitemShut {NoStop}%
\bibitem [{\citenamefont {Buzdin}(2008)}]{HelicalRef3}%
  \BibitemOpen
  \bibfield  {author} {\bibinfo {author} {\bibfnamefont {A.}~\bibnamefont {Buzdin}},\ }\bibfield  {title} {\bibinfo {title} {Direct coupling between magnetism and superconducting current in the {J}osephson ${\ensuremath{\varphi}}_{0}$ junction},\ }\href {https://doi.org/10.1103/PhysRevLett.101.107005} {\bibfield  {journal} {\bibinfo  {journal} {Phys. Rev. Lett.}\ }\textbf {\bibinfo {volume} {101}},\ \bibinfo {pages} {107005} (\bibinfo {year} {2008})}\BibitemShut {NoStop}%
\bibitem [{\citenamefont {Agterberg}(2012)}]{agterberg2012book}%
  \BibitemOpen
  \bibfield  {author} {\bibinfo {author} {\bibfnamefont {D.~F.}\ \bibnamefont {Agterberg}},\ }in\ \href {https://doi.org/10.1007/978-3-642-24624-1} {\emph {\bibinfo {booktitle} {Non-centrosymmetric superconductors}}},\ \bibinfo {series} {Lecture notes in physics}, Vol.\ \bibinfo {volume} {847},\ \bibinfo {editor} {edited by\ \bibinfo {editor} {\bibfnamefont {E.}~\bibnamefont {Bauer}}\ and\ \bibinfo {editor} {\bibfnamefont {M.}~\bibnamefont {Sigrist}}}\ (\bibinfo  {publisher} {Springer},\ \bibinfo {year} {2012})\ Chap.~\bibinfo {chapter} {5}\BibitemShut {NoStop}%
\bibitem [{\citenamefont {Mineev}\ and\ \citenamefont {Samokhin}(1994)}]{mineev1994}%
  \BibitemOpen
  \bibfield  {author} {\bibinfo {author} {\bibfnamefont {V.~P.}\ \bibnamefont {Mineev}}\ and\ \bibinfo {author} {\bibfnamefont {K.~V.}\ \bibnamefont {Samokhin}},\ }\bibfield  {title} {\bibinfo {title} {Helical phases in superconductor},\ }\href@noop {} {\bibfield  {journal} {\bibinfo  {journal} {Zh. Eksp. Teor. Fiz.}\ }\textbf {\bibinfo {volume} {105}},\ \bibinfo {pages} {747} (\bibinfo {year} {1994})},\ \translation{Sov. Phys. JETP 78, 401 (1994)}\BibitemShut {NoStop}%
\bibitem [{\citenamefont {Edelstein}(1996)}]{edelstein1996}%
  \BibitemOpen
  \bibfield  {author} {\bibinfo {author} {\bibfnamefont {V.~M.}\ \bibnamefont {Edelstein}},\ }\bibfield  {title} {\bibinfo {title} {The {Ginzburg}--{Landau} equation for superconductors of polar symmetry},\ }\href {https://doi.org/10.1088/0953-8984/8/3/012} {\bibfield  {journal} {\bibinfo  {journal} {J. Phys.: Condens. Matter}\ }\textbf {\bibinfo {volume} {8}},\ \bibinfo {pages} {339} (\bibinfo {year} {1996})}\BibitemShut {NoStop}%
\bibitem [{\citenamefont {Kitamura}\ \emph {et~al.}(2023)\citenamefont {Kitamura}, \citenamefont {Kanasugi}, \citenamefont {Chazono},\ and\ \citenamefont {Yanase}}]{QG_AnaSup}%
  \BibitemOpen
  \bibfield  {author} {\bibinfo {author} {\bibfnamefont {T.}~\bibnamefont {Kitamura}}, \bibinfo {author} {\bibfnamefont {S.}~\bibnamefont {Kanasugi}}, \bibinfo {author} {\bibfnamefont {M.}~\bibnamefont {Chazono}},\ and\ \bibinfo {author} {\bibfnamefont {Y.}~\bibnamefont {Yanase}},\ }\bibfield  {title} {\bibinfo {title} {Quantum geometry induced anapole superconductivity},\ }\href {https://doi.org/10.1103/PhysRevB.107.214513} {\bibfield  {journal} {\bibinfo  {journal} {Phys. Rev. B}\ }\textbf {\bibinfo {volume} {107}},\ \bibinfo {pages} {214513} (\bibinfo {year} {2023})}\BibitemShut {NoStop}%
\bibitem [{\citenamefont {Hu}\ \emph {et~al.}(2024)\citenamefont {Hu}, \citenamefont {Chen},\ and\ \citenamefont {Law}}]{QG_SDE}%
  \BibitemOpen
  \bibfield  {author} {\bibinfo {author} {\bibfnamefont {J.-X.}\ \bibnamefont {Hu}}, \bibinfo {author} {\bibfnamefont {S.~A.}\ \bibnamefont {Chen}},\ and\ \bibinfo {author} {\bibfnamefont {K.~T.}\ \bibnamefont {Law}},\ }\href {https://arxiv.org/abs/2403.01080} {\bibinfo {title} {Band-geometric origin of superconducting diode effect}} (\bibinfo {year} {2024}),\ \Eprint {https://arxiv.org/abs/2403.01080} {arXiv:2403.01080 [cond-mat.supr-con]} \BibitemShut {NoStop}%
\bibitem [{\citenamefont {Cao}\ \emph {et~al.}(2018)\citenamefont {Cao}, \citenamefont {Fatemi}, \citenamefont {Fang}, \citenamefont {Watanabe}, \citenamefont {Taniguchi}, \citenamefont {Kaxiras},\ and\ \citenamefont {Jarillo-Herrero}}]{OrigTBGSC}%
  \BibitemOpen
  \bibfield  {author} {\bibinfo {author} {\bibfnamefont {Y.}~\bibnamefont {Cao}}, \bibinfo {author} {\bibfnamefont {V.}~\bibnamefont {Fatemi}}, \bibinfo {author} {\bibfnamefont {S.}~\bibnamefont {Fang}}, \bibinfo {author} {\bibfnamefont {K.}~\bibnamefont {Watanabe}}, \bibinfo {author} {\bibfnamefont {T.}~\bibnamefont {Taniguchi}}, \bibinfo {author} {\bibfnamefont {E.}~\bibnamefont {Kaxiras}},\ and\ \bibinfo {author} {\bibfnamefont {P.}~\bibnamefont {Jarillo-Herrero}},\ }\bibfield  {title} {\bibinfo {title} {Unconventional superconductivity in magic-angle graphene superlattices},\ }\href {https://doi.org/10.1038/nature26160} {\bibfield  {journal} {\bibinfo  {journal} {Nature}\ }\textbf {\bibinfo {volume} {556}},\ \bibinfo {pages} {43–50} (\bibinfo {year} {2018})}\BibitemShut {NoStop}%
\bibitem [{\citenamefont {Zhou}\ \emph {et~al.}(2021)\citenamefont {Zhou}, \citenamefont {Xie}, \citenamefont {Taniguchi}, \citenamefont {Watanabe},\ and\ \citenamefont {Young}}]{RTriGSC}%
  \BibitemOpen
  \bibfield  {author} {\bibinfo {author} {\bibfnamefont {H.}~\bibnamefont {Zhou}}, \bibinfo {author} {\bibfnamefont {T.}~\bibnamefont {Xie}}, \bibinfo {author} {\bibfnamefont {T.}~\bibnamefont {Taniguchi}}, \bibinfo {author} {\bibfnamefont {K.}~\bibnamefont {Watanabe}},\ and\ \bibinfo {author} {\bibfnamefont {A.~F.}\ \bibnamefont {Young}},\ }\bibfield  {title} {\bibinfo {title} {Superconductivity in rhombohedral trilayer graphene},\ }\href {https://doi.org/10.1038/s41586-021-03926-0} {\bibfield  {journal} {\bibinfo  {journal} {Nature}\ }\textbf {\bibinfo {volume} {598}},\ \bibinfo {pages} {434} (\bibinfo {year} {2021})}\BibitemShut {NoStop}%
\bibitem [{\citenamefont {Han}\ \emph {et~al.}(2025)\citenamefont {Han}, \citenamefont {Lu}, \citenamefont {Hadjri}, \citenamefont {Shi}, \citenamefont {Wu}, \citenamefont {Xu}, \citenamefont {Yao}, \citenamefont {Cotten}, \citenamefont {Sedeh}, \citenamefont {Weldeyesus}, \citenamefont {Yang}, \citenamefont {Seo}, \citenamefont {Ye}, \citenamefont {Zhou}, \citenamefont {Liu}, \citenamefont {Shi}, \citenamefont {Hua}, \citenamefont {Watanabe}, \citenamefont {Taniguchi}, \citenamefont {Xiong}, \citenamefont {Zumbühl}, \citenamefont {Fu},\ and\ \citenamefont {Ju}}]{RTetraGSC1}%
  \BibitemOpen
  \bibfield  {author} {\bibinfo {author} {\bibfnamefont {T.}~\bibnamefont {Han}}, \bibinfo {author} {\bibfnamefont {Z.}~\bibnamefont {Lu}}, \bibinfo {author} {\bibfnamefont {Z.}~\bibnamefont {Hadjri}}, \bibinfo {author} {\bibfnamefont {L.}~\bibnamefont {Shi}}, \bibinfo {author} {\bibfnamefont {Z.}~\bibnamefont {Wu}}, \bibinfo {author} {\bibfnamefont {W.}~\bibnamefont {Xu}}, \bibinfo {author} {\bibfnamefont {Y.}~\bibnamefont {Yao}}, \bibinfo {author} {\bibfnamefont {A.~A.}\ \bibnamefont {Cotten}}, \bibinfo {author} {\bibfnamefont {O.~S.}\ \bibnamefont {Sedeh}}, \bibinfo {author} {\bibfnamefont {H.}~\bibnamefont {Weldeyesus}}, \bibinfo {author} {\bibfnamefont {J.}~\bibnamefont {Yang}}, \bibinfo {author} {\bibfnamefont {J.}~\bibnamefont {Seo}}, \bibinfo {author} {\bibfnamefont {S.}~\bibnamefont {Ye}}, \bibinfo {author} {\bibfnamefont {M.}~\bibnamefont {Zhou}}, \bibinfo {author} {\bibfnamefont {H.}~\bibnamefont {Liu}}, \bibinfo {author} {\bibfnamefont {G.}~\bibnamefont {Shi}}, \bibinfo {author} {\bibfnamefont
  {Z.}~\bibnamefont {Hua}}, \bibinfo {author} {\bibfnamefont {K.}~\bibnamefont {Watanabe}}, \bibinfo {author} {\bibfnamefont {T.}~\bibnamefont {Taniguchi}}, \bibinfo {author} {\bibfnamefont {P.}~\bibnamefont {Xiong}}, \bibinfo {author} {\bibfnamefont {D.~M.}\ \bibnamefont {Zumbühl}}, \bibinfo {author} {\bibfnamefont {L.}~\bibnamefont {Fu}},\ and\ \bibinfo {author} {\bibfnamefont {L.}~\bibnamefont {Ju}},\ }\href {https://arxiv.org/abs/2408.15233} {\bibinfo {title} {Signatures of chiral superconductivity in rhombohedral graphene}} (\bibinfo {year} {2025}),\ \Eprint {https://arxiv.org/abs/2408.15233} {arXiv:2408.15233 [cond-mat.mes-hall]} \BibitemShut {NoStop}%
\bibitem [{\citenamefont {Choi}\ \emph {et~al.}(2025)\citenamefont {Choi}, \citenamefont {Choi}, \citenamefont {Valentini}, \citenamefont {Patterson}, \citenamefont {Holleis}, \citenamefont {Sheekey}, \citenamefont {Stoyanov}, \citenamefont {Cheng}, \citenamefont {Taniguchi}, \citenamefont {Watanabe},\ and\ \citenamefont {Young}}]{RTetraGSC2}%
  \BibitemOpen
  \bibfield  {author} {\bibinfo {author} {\bibfnamefont {Y.}~\bibnamefont {Choi}}, \bibinfo {author} {\bibfnamefont {Y.}~\bibnamefont {Choi}}, \bibinfo {author} {\bibfnamefont {M.}~\bibnamefont {Valentini}}, \bibinfo {author} {\bibfnamefont {C.~L.}\ \bibnamefont {Patterson}}, \bibinfo {author} {\bibfnamefont {L.~F.}\ \bibnamefont {Holleis}}, \bibinfo {author} {\bibfnamefont {O.~I.}\ \bibnamefont {Sheekey}}, \bibinfo {author} {\bibfnamefont {H.}~\bibnamefont {Stoyanov}}, \bibinfo {author} {\bibfnamefont {X.}~\bibnamefont {Cheng}}, \bibinfo {author} {\bibfnamefont {T.}~\bibnamefont {Taniguchi}}, \bibinfo {author} {\bibfnamefont {K.}~\bibnamefont {Watanabe}},\ and\ \bibinfo {author} {\bibfnamefont {A.}~\bibnamefont {Young}},\ }\bibfield  {title} {\bibinfo {title} {Superconductivity and quantized anomalous hall effect in rhombohedral graphene},\ }\href@noop {} {\bibfield  {journal} {\bibinfo  {journal} {Nature}\ ,\ \bibinfo {pages} {1}} (\bibinfo {year} {2025})}\BibitemShut {NoStop}%
\bibitem [{\citenamefont {Diez-Merida}\ \emph {et~al.}(2023)\citenamefont {Diez-Merida}, \citenamefont {D{\'\i}ez-Carl{\'o}n}, \citenamefont {Yang}, \citenamefont {Xie}, \citenamefont {Gao}, \citenamefont {Senior}, \citenamefont {Watanabe}, \citenamefont {Taniguchi}, \citenamefont {Lu}, \citenamefont {Higginbotham} \emph {et~al.}}]{TBGSDE1}%
  \BibitemOpen
  \bibfield  {author} {\bibinfo {author} {\bibfnamefont {J.}~\bibnamefont {Diez-Merida}}, \bibinfo {author} {\bibfnamefont {A.}~\bibnamefont {D{\'\i}ez-Carl{\'o}n}}, \bibinfo {author} {\bibfnamefont {S.}~\bibnamefont {Yang}}, \bibinfo {author} {\bibfnamefont {Y.-M.}\ \bibnamefont {Xie}}, \bibinfo {author} {\bibfnamefont {X.-J.}\ \bibnamefont {Gao}}, \bibinfo {author} {\bibfnamefont {J.}~\bibnamefont {Senior}}, \bibinfo {author} {\bibfnamefont {K.}~\bibnamefont {Watanabe}}, \bibinfo {author} {\bibfnamefont {T.}~\bibnamefont {Taniguchi}}, \bibinfo {author} {\bibfnamefont {X.}~\bibnamefont {Lu}}, \bibinfo {author} {\bibfnamefont {A.~P.}\ \bibnamefont {Higginbotham}}, \emph {et~al.},\ }\bibfield  {title} {\bibinfo {title} {Symmetry-broken {J}osephson junctions and superconducting diodes in magic-angle twisted bilayer graphene},\ }\href@noop {} {\bibfield  {journal} {\bibinfo  {journal} {Nature Communications}\ }\textbf {\bibinfo {volume} {14}},\ \bibinfo {pages} {2396} (\bibinfo {year} {2023})}\BibitemShut
  {NoStop}%
\bibitem [{\citenamefont {Diez-Carlon}\ \emph {et~al.}(2025)\citenamefont {Diez-Carlon}, \citenamefont {Diez-Merida}, \citenamefont {Rout}, \citenamefont {Sedov}, \citenamefont {Virtanen}, \citenamefont {Banerjee}, \citenamefont {Penttila}, \citenamefont {Altpeter}, \citenamefont {Watanabe}, \citenamefont {Taniguchi}, \citenamefont {Yang}, \citenamefont {Law}, \citenamefont {Heikkil\"a}, \citenamefont {T\"orm\"a}, \citenamefont {Scheurer},\ and\ \citenamefont {Efetov}}]{TBGSDE2}%
  \BibitemOpen
  \bibfield  {author} {\bibinfo {author} {\bibfnamefont {A.}~\bibnamefont {Diez-Carlon}}, \bibinfo {author} {\bibfnamefont {J.}~\bibnamefont {Diez-Merida}}, \bibinfo {author} {\bibfnamefont {P.}~\bibnamefont {Rout}}, \bibinfo {author} {\bibfnamefont {D.}~\bibnamefont {Sedov}}, \bibinfo {author} {\bibfnamefont {P.}~\bibnamefont {Virtanen}}, \bibinfo {author} {\bibfnamefont {S.}~\bibnamefont {Banerjee}}, \bibinfo {author} {\bibfnamefont {R.~P.~S.}\ \bibnamefont {Penttila}}, \bibinfo {author} {\bibfnamefont {P.}~\bibnamefont {Altpeter}}, \bibinfo {author} {\bibfnamefont {K.}~\bibnamefont {Watanabe}}, \bibinfo {author} {\bibfnamefont {T.}~\bibnamefont {Taniguchi}}, \bibinfo {author} {\bibfnamefont {S.~Y.}\ \bibnamefont {Yang}}, \bibinfo {author} {\bibfnamefont {K.~T.}\ \bibnamefont {Law}}, \bibinfo {author} {\bibfnamefont {T.~T.}\ \bibnamefont {Heikkil\"a}}, \bibinfo {author} {\bibfnamefont {P.}~\bibnamefont {T\"orm\"a}}, \bibinfo {author} {\bibfnamefont {M.~S.}\ \bibnamefont {Scheurer}},\ and\ \bibinfo {author}
  {\bibfnamefont {D.~K.}\ \bibnamefont {Efetov}},\ }\href {https://arxiv.org/abs/2502.04785} {\bibinfo {title} {Probing the flat-band limit of the superconducting proximity effect in twisted bilayer graphene {J}osephson junctions}} (\bibinfo {year} {2025}),\ \Eprint {https://arxiv.org/abs/2502.04785} {arXiv:2502.04785 [cond-mat.supr-con]} \BibitemShut {NoStop}%
\bibitem [{\citenamefont {Lin}\ \emph {et~al.}(2022)\citenamefont {Lin}, \citenamefont {Siriviboon}, \citenamefont {Scammell}, \citenamefont {Liu}, \citenamefont {Rhodes}, \citenamefont {Watanabe}, \citenamefont {Taniguchi}, \citenamefont {Hone}, \citenamefont {Scheurer},\ and\ \citenamefont {Li}}]{TTGSDE}%
  \BibitemOpen
  \bibfield  {author} {\bibinfo {author} {\bibfnamefont {J.-X.}\ \bibnamefont {Lin}}, \bibinfo {author} {\bibfnamefont {P.}~\bibnamefont {Siriviboon}}, \bibinfo {author} {\bibfnamefont {H.~D.}\ \bibnamefont {Scammell}}, \bibinfo {author} {\bibfnamefont {S.}~\bibnamefont {Liu}}, \bibinfo {author} {\bibfnamefont {D.}~\bibnamefont {Rhodes}}, \bibinfo {author} {\bibfnamefont {K.}~\bibnamefont {Watanabe}}, \bibinfo {author} {\bibfnamefont {T.}~\bibnamefont {Taniguchi}}, \bibinfo {author} {\bibfnamefont {J.}~\bibnamefont {Hone}}, \bibinfo {author} {\bibfnamefont {M.~S.}\ \bibnamefont {Scheurer}},\ and\ \bibinfo {author} {\bibfnamefont {J.}~\bibnamefont {Li}},\ }\bibfield  {title} {\bibinfo {title} {Zero-field superconducting diode effect in small-twist-angle trilayer graphene},\ }\href {https://doi.org/10.1038/s41567-022-01700-1} {\bibfield  {journal} {\bibinfo  {journal} {Nature Physics}\ }\textbf {\bibinfo {volume} {18}},\ \bibinfo {pages} {1221} (\bibinfo {year} {2022})}\BibitemShut {NoStop}%
\bibitem [{\citenamefont {Han}\ \emph {et~al.}(2024)\citenamefont {Han}, \citenamefont {Herzog-Arbeitman}, \citenamefont {Bernevig},\ and\ \citenamefont {Kivelson}}]{QGNest}%
  \BibitemOpen
  \bibfield  {author} {\bibinfo {author} {\bibfnamefont {Z.}~\bibnamefont {Han}}, \bibinfo {author} {\bibfnamefont {J.}~\bibnamefont {Herzog-Arbeitman}}, \bibinfo {author} {\bibfnamefont {B.~A.}\ \bibnamefont {Bernevig}},\ and\ \bibinfo {author} {\bibfnamefont {S.~A.}\ \bibnamefont {Kivelson}},\ }\bibfield  {title} {\bibinfo {title} {``{Q}uantum geometric nesting" and solvable model flat-band systems},\ }\bibfield  {journal} {\bibinfo  {journal} {Physical Review X}\ }\textbf {\bibinfo {volume} {14}},\ \href {https://doi.org/10.1103/physrevx.14.041004} {10.1103/physrevx.14.041004} (\bibinfo {year} {2024})\BibitemShut {NoStop}%
\bibitem [{\citenamefont {Hu}\ \emph {et~al.}(2023)\citenamefont {Hu}, \citenamefont {Sun}, \citenamefont {Xie},\ and\ \citenamefont {Law}}]{PhysRevLett.130.266003}%
  \BibitemOpen
  \bibfield  {author} {\bibinfo {author} {\bibfnamefont {J.-X.}\ \bibnamefont {Hu}}, \bibinfo {author} {\bibfnamefont {Z.-T.}\ \bibnamefont {Sun}}, \bibinfo {author} {\bibfnamefont {Y.-M.}\ \bibnamefont {Xie}},\ and\ \bibinfo {author} {\bibfnamefont {K.~T.}\ \bibnamefont {Law}},\ }\bibfield  {title} {\bibinfo {title} {Josephson diode effect induced by valley polarization in twisted bilayer graphene},\ }\href {https://doi.org/10.1103/PhysRevLett.130.266003} {\bibfield  {journal} {\bibinfo  {journal} {Phys. Rev. Lett.}\ }\textbf {\bibinfo {volume} {130}},\ \bibinfo {pages} {266003} (\bibinfo {year} {2023})}\BibitemShut {NoStop}%
\bibitem [{\citenamefont {Dunbrack}(2026)}]{CodeRef}%
  \BibitemOpen
  \bibfield  {author} {\bibinfo {author} {\bibfnamefont {A.}~\bibnamefont {Dunbrack}},\ }\href {https://doi.org/10.5281/zenodo.19731376} {\bibinfo {title} {Quantum geometric helical {SC} wavevector in 1d bipartite lattice with {IT} symmetry}} (\bibinfo {year} {2026})\BibitemShut {NoStop}%
\bibitem [{\citenamefont {Cheng}(2013)}]{QGT_PedRef}%
  \BibitemOpen
  \bibfield  {author} {\bibinfo {author} {\bibfnamefont {R.}~\bibnamefont {Cheng}},\ }\href {https://arxiv.org/abs/1012.1337} {\bibinfo {title} {Quantum geometric tensor ({F}ubini-{S}tudy metric) in simple quantum system: A pedagogical introduction}} (\bibinfo {year} {2013}),\ \Eprint {https://arxiv.org/abs/1012.1337} {arXiv:1012.1337 [quant-ph]} \BibitemShut {NoStop}%
\bibitem [{\citenamefont {Avdoshkin}\ and\ \citenamefont {Popov}(2023)}]{HigherorderBargmann}%
  \BibitemOpen
  \bibfield  {author} {\bibinfo {author} {\bibfnamefont {A.}~\bibnamefont {Avdoshkin}}\ and\ \bibinfo {author} {\bibfnamefont {F.~K.}\ \bibnamefont {Popov}},\ }\bibfield  {title} {\bibinfo {title} {Extrinsic geometry of quantum states},\ }\href {https://doi.org/10.1103/PhysRevB.107.245136} {\bibfield  {journal} {\bibinfo  {journal} {Phys. Rev. B}\ }\textbf {\bibinfo {volume} {107}},\ \bibinfo {pages} {245136} (\bibinfo {year} {2023})}\BibitemShut {NoStop}%
\bibitem [{\citenamefont {Ma}\ \emph {et~al.}(2010)\citenamefont {Ma}, \citenamefont {Chen}, \citenamefont {Fan},\ and\ \citenamefont {Liu}}]{NonAbelQGT}%
  \BibitemOpen
  \bibfield  {author} {\bibinfo {author} {\bibfnamefont {Y.-Q.}\ \bibnamefont {Ma}}, \bibinfo {author} {\bibfnamefont {S.}~\bibnamefont {Chen}}, \bibinfo {author} {\bibfnamefont {H.}~\bibnamefont {Fan}},\ and\ \bibinfo {author} {\bibfnamefont {W.-M.}\ \bibnamefont {Liu}},\ }\bibfield  {title} {\bibinfo {title} {Abelian and non-{A}belian quantum geometric tensor},\ }\href {https://doi.org/10.1103/PhysRevB.81.245129} {\bibfield  {journal} {\bibinfo  {journal} {Phys. Rev. B}\ }\textbf {\bibinfo {volume} {81}},\ \bibinfo {pages} {245129} (\bibinfo {year} {2010})}\BibitemShut {NoStop}%
\bibitem [{\citenamefont {Schulz}(1994)}]{CompareCDWandSC}%
  \BibitemOpen
  \bibfield  {author} {\bibinfo {author} {\bibfnamefont {H.~J.}\ \bibnamefont {Schulz}},\ }\href {https://arxiv.org/abs/cond-mat/9402103} {\bibinfo {title} {Functional integrals for correlated electrons}} (\bibinfo {year} {1994}),\ \Eprint {https://arxiv.org/abs/cond-mat/9402103} {arXiv:cond-mat/9402103 [cond-mat]} \BibitemShut {NoStop}%
\end{thebibliography}%

\appendix
\section{Overview of quantum geometry}\label{Apx:QGReview}

Broadly speaking, \textit{quantum geometry} describes how the eigen\textit{vectors} of some family of Hamiltonians varies as one varies some continuous parameter. In condensed matter physics, this parameter is traditionally the crystal quasimomentum $k$, but it is also used in quantum information and cold-atom systems with the parameter being some adiabatically tunable knob (see, e.g., Ref. \cite{QGT_PedRef}). The latter usage is in line with the typical introduction of Berry curvature in a graduate quantum mechanics course.

A subtle point is that the eigenvectors themselves are only determined up to phase (and up to unitary transformation when they are energy-degenerate). Accordingly, physical quantum geometric quantities must be invariant under these phase multiplications (generally gauge transformations). The usual method for avoiding such complications at a computational level is to express all quantities in terms of traces of projectors onto different states/bands.

This quantum geometry can be described completely \cite{HigherorderBargmann} by the trace of triplets of projectors: if the energies $E_n(x)$ and the traces $\Tr[P_n(x)P_m(y)P_l(z)]$ are known for all $x$, $y$, and $z$ in the parameter space and $n$, $m$, and $l$ indexing the eigenvalues of $H$, then one can reconstruct the full Hamiltonian $H(x)$ up to a global unitary change of basis.

For simplicity, suppose now we are interested in only one band, allowing us to drop the $nml$ indices and just write $\Tr[P(x)P(y)P(z)]$. The quantities that arise most commonly in physical applications arise from the lowest nontrivial contributions to the Taylor expansion near $x=y=z$:
\begin{equation}\begin{split}
    &\Tr[P(x)P(x+\delta)P(x+\epsilon)]=\\
    &\qquad\quad 1+\delta_\mu\epsilon_\nu\Tr[P(\partial^\mu P)(\partial^\nu P)]/2+O(\delta^2,\epsilon^2) 
\end{split}\end{equation}

This trace defines the \textit{quantum geometric tensor},
\begin{equation}
    \QGT^{\mu\nu}=2\Tr[P(\partial^\mu P)(\partial^\nu P)],
\end{equation}
a Hermitian matrix. The (symmetric) real part is called the \textit{quantum metric},
\begin{equation}
    g^{\mu\nu}=2\Re[\Tr[P(\partial^\mu P)(\partial^\nu P)]]=\Tr[(\partial^\mu P)(\partial^\nu P)],
\end{equation}
and is equal to the pullback via the map $P(k)$ of the Frobenius inner product. The (antisymmetric) imaginary part of the quantum geometric tensor is the \textit{Berry curvature},
\begin{equation}
    \Omega^{\mu\nu}=4\Im[\Tr[P(\partial^\mu P)(\partial^\nu P)]],
\end{equation}
which in 2D is usually reduced from a 2-form to a pseudoscalar by contraction with the Levi-Civita tensor. The integral of this pseudoscalar is $2\pi$ times the Chern number, a topological invariant.

These expressions then naturally lead to their derivatives, $\partial_\mu g_{\nu\rho}$ and $\partial_\mu\Omega_{\nu\rho}$, called the \textit{quantum metric dipole} and \textit{Berry curvature dipole}, as well as higher moments. There are also higher-order terms in the expansion of the trace of projectors, most of which cannot be expressed in terms of the quantum geometric tensor \cite{HigherorderBargmann} but can still be relevant to physical quantities, including those discussed in this paper.

Finally, rather than considering only one particular band, one could consider the band indices $nml$ as well. Preserving these indices generates the study of \textit{non-Abelian quantum geometry}: the \textit{non-Abelian quantum geometric tensor} \cite{NonAbelQGT} of some set of bands with projector $P$ is defined by
\begin{equation}
    \QGT^{\mu\nu}_{mn}=2\Tr[P(\partial^\mu P_m)(\partial^\nu P_n)],
\end{equation}
with $m,n$ running over band indices.

\section{Derivation of \cref{eq:FreeEnResult}}\label{Apx:FreeEnResult}
We here provide the full derivation of \cref{eq:FreeEnResult}. We use the notation that $ij$ run over orbitals within a unit cell (including different sublattices), whereas capital $IJ$ to refer to the orbitals of individual lattice sites.

The GL free energy is
\begin{equation}
    \Omega=U^{-1}\Tr[\hat\Delta^\dagger\hat\Delta]-T\sum_\omega\Tr[\ln(i\omega-H)].
\end{equation}
Here, we assume a local interaction $U$, and $\hat{\Delta}_{IJ}\propto\delta_{IJ}$ is a diagonal matrix. The BdG Hamiltonian $H$ is 
\begin{equation}
    H=\begin{bmatrix}H_0&\hat\Delta\\\hat\Delta^\dagger&-H_0^T\end{bmatrix},
\end{equation}
and the trace $\Tr$ should be understood in real space as running over $IJ$ indices.
(The chemical potential $\mu$ has been absorbed into $H_0$.)

Expanding the trace log in $\Delta$ and considering only the quadratic term in $\Delta$ gives the contribution
\begin{equation}\begin{split}
    \Omega^{(2)}=&U^{-1}\Tr[\hat\Delta^\dagger\hat\Delta]\\&+\frac{T}{2}\sum_\omega\Tr[(i\omega-H_0)^{-1}\hat\Delta (i\omega+H_0^T)^{-1}\hat\Delta^\dagger].
\end{split}\end{equation}
This order of expansion is sufficient for studying the lowest-order gradient terms, the Lifshitz invariants and the superfluid weight.

We now make a series of assertions about the nature of superconductivity to reduce to the case where expressions in terms of quantum geometry are possible.

First, we assume spin singlet superconductivity and write $\hat\Delta=(i\sigma_y)\hat\Delta_{\sigma_0}$. Using the relation
\begin{equation}
    i\sigma_y(i\omega+H_0^T)(-i\sigma_y)=-\Ts (i\omega-H_0) \Ts^{-1}
\end{equation}
and using the Green's function $G=(i\omega-H_0)^{-1}$, we can now write

\begin{equation}\begin{split}
    \Omega^{(2)}=&U^{-1}\Tr[\hat\Delta_{\sigma_0}^\dagger\hat\Delta_{\sigma_0}]\\
    &-\frac{T}{2}\sum_\omega\Tr[G\hat\Delta_{\sigma_0} \Ts G\Ts^{-1}\hat\Delta_{\sigma_0}^\dagger].
\end{split}\end{equation}

Assuming lattice translation symmetry, we now Fourier transform this expression to yield
\begin{equation}\begin{split}
    \hat\Delta_{\sigma_0,IJ}&=\sum_q \hat\Delta_{ij}(q)\delta_{r_I,r_J}e^{iq\cdot r_I}\\
    G_{IJ}&=\sum_q G_{ij}(q)e^{iq\cdot (r_I-r_J)},
\end{split}\end{equation}
whereupon
\begin{equation}\begin{split}
    \Omega^{(2)}=&\sum_q U^{-1}\Tr[\hat \Delta(q)\hat\Delta(q)^\dagger]\\
    &-\frac{T}{2}\sum_{\omega,k,q}\Tr[G(k)\hat\Delta(q)\Ts G(-k-q)\Ts^{-1}\hat\Delta(q)^\dagger].
\end{split}\end{equation}
Now the traces should be understood as running over the indices $ij$ only.

Next, we express $G(k)$ in terms of projectors $P_{k,n}$ and energies $\epsilon_n(k)$ onto band $n$ of $H_0$ as
\begin{equation}
    G(k)=\sum_n\frac{P_{k,n}}{i\omega-\epsilon_n(k)}
\end{equation}
implying
\begin{equation}\begin{split}
    \Omega^{(2)}=&\sum_q U^{-1}\Tr[\hat \Delta(q)\hat\Delta(q)^\dagger]\\
    &-\frac{T}{2}\sum_{\omega,k,q,m,n}\frac{\Tr[P_{k,m}\hat\Delta(q)\Ts P_{-k-q,n}\Ts^{-1}\hat\Delta(q)^\dagger]}{(i\omega-\epsilon_m(k))(i\omega-\epsilon_n(-k-q))}.
\end{split}\end{equation}
At this stage, we assert the separation of the spectrum into low and high-energy bands discussed in the main text and thereby approximate the expression by evaluating the $m,n$ sums only over low-energy bands $\bar m,\bar n$, producing the low-energy Green's function
\begin{equation}
    \bar G=\sum_{\bar n}\frac{P_{k,\bar n}}{i\omega-\epsilon_{\bar n}(k)}
\end{equation}

Next, we assert the \textit{uniform pairing condition} \cite{OrigQMSFWeight}, which states that the pairing is the same magnitude in all low-energy bands. Algebraically, within the low-energy bands, this means
\begin{equation}
    \hat\Delta(q)=\Delta_0(q)V(q)
\end{equation}
where $\Delta_0$ is a scalar representing the magnitude of pairing and $V$ is some diagonal unitary matrix acting on the orbital basis.

The free energy therefore simplifies to

\begin{equation}\begin{split}\label{eq:apx-preminQM}
    \Omega^{(2)}=\sum_q&|\Delta_0(q)|^2\Big\{U^{-1}N\\
    &-\frac{T}{2}\sum_{\omega,k}\Tr[\bar G(k)V(q)\Ts \bar G(-k-q)\Ts^{-1}V(q)^\dagger]\Big\},
\end{split}\end{equation}
where $N$ is the number of bands.

For small $q$ (and, for the time-reversal-symmetry breaking case discussed in the main text, small $\alpha$), $V(q)$ can be eliminated with a suitable choice of location for the origins of the orbitals, giving the \textit{minimal quantum metric}. This is discussed at length in \cref{Apx:MinQuantMet}, so we presume this has been done and accordingly reduce the expression for the free energy to
\begin{equation}\begin{split}\label{eq:FreeEnNonDegenBands}
    \Omega^{(2)}=\sum_q&|\Delta_0(q)|^2\Big\{U^{-1}N\\
    &-\frac{T}{2}\sum_{\omega,k}\Tr[\bar G(k)\Ts \bar G(-k-q)\Ts^{-1}]\Big\}\\
    =\sum_q&|\Delta_0(q)|^2\Bigg\{U^{-1}N\\
    &-\frac{T}{2}\sum_{\omega,k,\bar m,\bar n}\frac{\Tr[P_{k,\bar m}\Ts P_{-k-q,\bar n}\Ts^{-1}]}{(i\omega-\epsilon_{\bar m}(k))(-i\omega-\epsilon_{\bar n}(-k-q))}\Bigg\}.
\end{split}\end{equation}

In the presence of time-reversal symmetry, the expression in the numerator describes non-Abelian quantum geometry, and an expansion in $q$ reveals the quadratic term to be the non-Abelian quantum metric. In the absence of time-reversal symmetry, this can still be generalized to the augmented non-Abelian quantum metric in the same way.

We now simplify our results by taking the low-energy bands to be degenerate, such that $\epsilon_{\bar m}(k)=\epsilon(k)$. This would naturally follow if the model has its low-energy bands related by symmetry, which is also the most natural way to impose the uniform pairing condition.

Then, the sum over $\bar m,\bar n$ can be performed explicitly by summing the projectors onto individual bands into the projector onto all low-energy bands $P_k$, therefore reducing the non-Abelian quantum metric to the Abelian one. This produces the resulting free energy
\begin{equation}\begin{split}\label{eq:SCFEincdisp}
    \Omega^{(2)}=\sum_q&|\Delta_0(q)|^2\Big\{U^{-1}N\\
    &-\frac{T}{2}\sum_{\omega,k}\frac{\Tr[P_k\Ts P_{-k-q}\Ts^{-1}]}{(i\omega-\epsilon(k))(-i\omega-\epsilon(-k-q))}\Bigg\}.
\end{split}\end{equation}
In the flat-band limit where $\epsilon(k)=0$, the Matsubara sum can be evaluated explicitly as $\sum_\omega\frac{T}{\omega^2}=\frac{1}{4T}$, whereupon the expression simplifies to
\begin{equation}
    \Omega^{(2)}=\sum_q|\Delta_0(q)|^2\Big\{\frac{N}{U}-\frac{\sum_{k}\Tr[P_k\Ts P_{-k-q}\Ts^{-1}]}{8T}\Bigg\}
\end{equation}
as given in the main text in \cref{eq:FreeEnResult}.

\section{Minimal Quantum Metric}\label{Apx:MinQuantMet}
The $V(q)$ are free parameters, and $\Omega^{(2)}$ must be minimized with respect to this choice. However, it turns out that this minimization is perfectly equivalent to a different minimization problem that can be expressed purely in terms of quantum geometry, known as finding the minimal quantum metric. This fact was first identified in Ref.~\onlinecite{MinQuantMetric}.

We begin, for simplicity, from \cref{eq:apx-preminQM} and consider the simplest case of time-reversal-symmetric energy-degenerate bands. In this case, the interesting part of the calculation is the trace $\Tr[P_k V(q)P_{k+q}V(q)^\dagger]$, which contributes to the free energy with a minus sign and so must be maximized.

Consider a particular $q=q_0$ and select a vector of Hermitian diagonal matrices $M$ such that $V(q_0)=e^{iq_0\cdot M}$, which is always possible to do. Then, write $\tilde V(q)=e^{iq\cdot M}$ for that choice of $M$. If we then define
\begin{equation}
    \tilde P_k=\tilde V(k)P_k\tilde V(k)^\dagger
\end{equation}
then
\begin{equation}
\Tr[P_kV(q_0)P_{k+q_0}V(q_0)^\dagger]=\Tr[\tilde P_k\tilde P_{k+q_0}]
\end{equation}
Therefore, including maximization of the left-hand side trace over all $V(q)$ is equivalent to maximizing the right-hand side trace over all $\tilde P_k$, i.e., over all changes of basis of the projectors of the form $e^{ik\cdot M}$.
This can, in turn, be related to the locations of orbital centers, as shown in Ref.~\onlinecite{MinQuantMetric}.

To quadratic order, this trace is $1-g_{ij}q^iq^j$. Therefore, for small $q$, one wants to choose a basis that minimizes the $q$-direction eigenvalue of $g_{ij}$. As it turns out, it is possible to do this for every direction in $k$-space simultaneously by minimizing the trace of $g_{ij}$, producing what is called the \textit{minimal quantum metric} \cite{MinQuantMetric}.

Now, we lift the constraint of time-reversal symmetry. As discussed in the main text, at the perturbative level, one can view the breaking of time-reversal symmetry as adding a third, continuous parameter $\alpha$ to $k$-space, whereupon the analogous trace is proportional to $1-g_{\mu\nu}q^\mu q^\nu$, with $\mu,\nu\in\{x,y,\alpha\}$. Since the situation is entirely analogous once reduced to this extended quantum metric, one can also assert that the appropriate choice of basis for the projectors $\tilde P$ (which can now vary not only with $k$ but also with $\alpha$) is the one which minimizes the trace of the extended quantum metric $g_{\mu\nu}$.

\section{Quantum geometry in dispersive bands}\label{Apx:QGDisp}
In this appendix, we consider the contributions of both quantum geometry and dispersion. We retain the assumptions of the main text (cf. \cref{fig:bandreqs}) other than restriction to exactly flat bands.

In the dispersive case (with degenerate bands), per \cref{eq:SCFEincdisp}, to quadratic order in $\Delta$, the free energy is
\begin{equation}\begin{split}
    \Omega^{(2)}=\sum_q&|\Delta_0(q)|^2\Big\{U^{-1}N\\
    &-\frac{T}{2}\sum_{\omega,k}\frac{\Tr[P_k\Ts P_{-k-q}\Ts^{-1}]}{(i\omega-\epsilon(k))(-i\omega-\epsilon(-k-q))}\Bigg\}.
\end{split}\end{equation}

In the main text where $\epsilon(k)=0$, only the projector trace was expanded in powers of $q$ (and $\alpha$). In the dispersive case, the denominator must be expanded as well. Since the degeneracy of the bands (cf. \cref{eq:FreeEnNonDegenBands}) enables the second term to be factored (for each $k$) into the purely quantum-geometric projector trace and the purely dispersive Matsubara sum, these expansions can be performed separately and multiplied, i.e.,
\begin{equation}\begin{split}
    &\Omega^{(2)}\sim |\Delta(q)|^2\Big\{U^{-1}N-\\
    \sum_k (a_0^g+d^g_iq^i+&D^{g}_{ij}q^iq^j+\ldots)(a_0^\epsilon+d^\epsilon_iq^i+D^\epsilon_{ij}q^iq^j+\ldots)\Big\}
\end{split}\end{equation}
where $a_0$, $d$, and $D$ are the contributions from the geometric ($g$) and the dispersive ($\epsilon$) parts separately.

\subsection{Combining quantum geometric and dispersive contributions}\label{Apx:MixedSFWeight}
Considering the individual coefficients shows that the Lifshitz invariant essentially additive,
\begin{equation}
    d_i\sim a_0^\epsilon d^g_i+a_0^gd^\epsilon_i,
\end{equation}
whereas the superfluid weight in general is not,
\begin{equation}
    d_{ij}\sim a_0^\epsilon D^g_{ij}+a_0^gD^\epsilon_{ij}+(d^g_id^\epsilon_j+d^\epsilon_id^g_j)/2.
\end{equation}
However, in the presence of time-reversal symmetry, the third term vanishes, giving the usual additivity of the geometric and dispersive contributions to superfluid weight \cite{OrigQMSFWeight}.

Since $d^g$ measures time-reversal symmetry breaking in the quantum geometry and $d^\epsilon$ measures time-reversal symmetry breaking in the dispersion, the new cross-term that depends on their product demands both, and is quadratic in time-reversal symmetry breaking overall. Note these expressions are before integrating over $k$, so the cross-term in the superfluid weight can be nonzero even in the presence of inversion symmetry (where the Lifshitz invariant must vanish) - only time-reversal symmetry breaking is required. 

This new cross-term in the superfluid weight is not necessarily positive-definite in itself: if a Hamiltonian $H$ causes this cross-term a positive contribution at some $q$, then the Hamiltonian $-H$ will change the sign of $d^\epsilon$ but not $d^g$ and therefore give a negative contribution. However, this will be dominated by the other contributions in the perturbative limit where $\alpha\ll g_{\alpha\alpha}^{-1/2}$ and $\alpha\ll \left|\frac{d\epsilon}{dk}\right| \left|\frac{d^2\epsilon}{d\alpha dk}\right|^{-1}$ for all $k$, and therefore will not make a pair density wave state unless $\alpha$ is large, at which point higher-order-in-$\alpha$ terms should be included.

\subsection{Comparing temperature dependence of quantum geometric and dispersive contributions}\label{Apx:TempDep}
Now we focus on a mostly flat band (i.e., working to lowest order in $\epsilon$), and compare the temperature dependence of quantum-geometric and dispersive contributions to the Lifshitz invariant and superfluid weight, and the implications for the helical wavevector. As this analysis is performed within Ginzburg-Landau theory, it is only valid near $T_c$. Accordingly, the most experimentally-relevant predictions are for properties which are temperature-independent near $T_c$, as all other relations are approximately linear and therefore indistinguishable from each other without specific knowledge of the coefficients.

In the quantum geometric part, the $q$ derivatives are taken of the trace. Then, the Matsubara sum reduces to $T\sum\omega^{-2}\sim T^{-1}$ both for the Lifshitz invariant and the superfluid weight. This implies that their ratio, the helical wavevector, is temperature-independent.

On the other hand, in the dispersive part, the $q$-derivatives are taken of the $\epsilon(-k-q)$ in the denominator. These derivatives bring extra powers of Matsubara frequency, which (sending $\epsilon\rightarrow 0$, but keeping $v=d\epsilon/dk$) gives a contribution proportional to $v\sum\omega^{-3}$ for the Lifshitz invariant. For the superfluid weight, there are two contributions scaling as $(dv/dk)T\sum\omega^{-3}$ and $v^2T\sum\omega^{-4}$ for superfluid weight; on the Fermi surface the latter is usually more important, but in this nearly-flat-band case the former will generally dominate due to being only linear in dispersion.

Hence, the dispersive Lifshitz invariant scales as $T^{-2}$, and the dispersive superfluid weight scales as $T^{-2}$ in the flat-band case and $T^{-3}$ in the Fermi surface case. This means that for a Fermi surface superconductor, the helical wavevector will scale as $T^{-1}$. For a nearly-flat-band superconductor where both the quantum geometry and dispersion are nontrivial, both the Lifshitz invariant and superfluid weight depend on a combination of factors scaling as $T^{-2}$ and $T^{-3}$, producing more complicated behavior of $q_0$, but if either one dominates both the Lifshitz and superfluid weight contributions, then one will have an approximately temperature-independent $q_0$.

\section{Changes in Critical Temperature}\label{Apx:TC}
We now consider the implications of time-reversal symmetry breaking on critical temperature, which are described by the $g_{\alpha\alpha}$ term of \cref{eq:QMinprojnoTRS}. 

Consider first a flat band with inversion symmetry, ensuring the helical wavevector vanishes ($q_0=0$). Then, \cref{eq:FreeEnResult} reduces to
\begin{equation}\label{eq:invsymmFmin}
    \Omega^{(2)}_{q=0}=|\Delta_0(0)|^2\left\{\frac{N}{U}-\frac{\sum_{k}\Tr[P_{k,\bar\alpha}P_{k,-\bar\alpha}]}{8T}\right\},
\end{equation}
giving a critical temperature
\begin{equation}
    T_c=\frac{U}{8N}\sum_{k}\Tr[P_{k,\bar\alpha}P_{k,-\bar\alpha}].
\end{equation}
To quadratic order in $\bar\alpha$, the critical temperature the projector trace is $T_c=T_{c,0}(1-2g_{\alpha\alpha}\bar\alpha^2)$. However, note the projector trace is also positive-definite, which implies the superconducting state always has some region of stability - there is no critical value of $\alpha$ above which there is no superconducting state, despite the vanishing of $T_c$ in the quadratic approximation.

In the absence of inversion symmetry, if a Lifshitz invariant is present, then \cref{eq:invsymmFmin} should be replaced with the evaluation of \cref{eq:FreeEnResult} at the new free-energy minimum, $q=q_0$, instead of $q=0$. The derivation of the change in $T_c$ is otherwise identical, so the quadratic-in-$\alpha$ estimate is
\begin{equation}
    T_c=T_{c,0}(1-2g_{\mu\nu}q_0^\mu q_0^\nu)
\end{equation}
where $q_0^\mu=(q_0^i,\bar\alpha)$ with $q_0^i$ the helical wavevector.

\section{Independent Perturbations \label{Apx:IndepPert}}
We here prove a sufficient condition for the quantum geometric Lifshitz invariant to vanish: if the perturbation $\alpha$ acts only on degrees of freedom that do not vary with $k$, then the off-diagonal quantum metric $g_{i\alpha}$, and hence the quantum geometric Lifshitz invariant, will vanish. More concretely, suppose the state space for each value of $k$ and $\alpha$ decomposes (independently of $k$ or $\alpha$) as $\mathcal{H}=\mathcal{H}_1\otimes\mathcal{H}_2$, and that the projectors $P(k,\alpha)$ can be decomposed as
\begin{equation}
    P(k,\alpha)=P_1(k)\otimes P_2(\alpha)
\end{equation}
where $P_i$ acts on the subspace $\mathcal{H}_i$. Then, because the trace of a tensor product is the product of traces,
\begin{equation}
    g_{i\alpha}=\Tr[(\partial_i P)(\partial_\alpha P)]=\Tr[(\partial_i P_1)P_1]\Tr[P_2(\partial_\alpha P_2)]=0.
\end{equation}

\section{Pseudo-time-reversal example}\label{Apx:PseudoTR}
Suppose the projectors of some time-reversal-symmetry-breaking superconductor instead obey the pseudo-time-reversal constraint $\Ts P_{-k}\Ts^{-1}=P_{k+2Q}$ for some fixed $Q$. From this constraint, one expects a helical wavevector of $2Q$ due to perfect quantum geometric nesting, as defined in \cite{QGNest}. We now show that this result also follows from the perturbative treatment.

This system is a perturbation of a time-reversal-symmetric system where $Q$ is a perturbative parameter. Define $P_0(k)=P(k+Q)$, which respects time-reversal symmetry,
\begin{equation}\begin{split}
    \Ts P_0(-k)\Ts^{-1}&=\Ts P(-k+Q)\Ts^{-1}=P(k-Q+2Q)\\&=P(k+Q)=P_0(k).
\end{split}\end{equation}
Then, take $P_{k,\alpha}=P_0(k-\alpha Q)$, such that $P_{k,0}=P_0(k)$, $P_{k,1}=P(k)$, and $\Ts P_{k,\alpha}\Ts^{-1}=P_{-k,-\alpha}$. The key identity that allows for a straightforward computation of the helical wavevector within perturbation theory is
\begin{equation}
    \partial_\alpha P_{k,\alpha}=-Q\cdot\partial_k P_{k,Q}.
\end{equation}

Consider for simplicity a coordinate system where $Q=|Q|\hat x$. Up to a factor of $-|Q|$, by the above identity, $g_{i\alpha}=g_{ix}$, so $\int g_{i\alpha}=\int g_{ix}$. Consequently, $[\int g_{ij}]^{-1}[\int g_{i\alpha}]=-Q$, and so since $\bar\alpha=1$, as expected, we have $q_0=2Q$ by \cref{eq:helicalwavevector}.

\section{Concrete bipartite lattice example}\label{Apx:BipartiteEx}
We now consider a specific lattice example described by the Hamiltonian in \cref{eq:genbipartiteham},
\begin{equation}
    H(k)=\begin{bmatrix}
        0&f(k)&f^*(k)\\
        f^*(k)&0&0\\
        f(k)&0&0
    \end{bmatrix},
\end{equation}
corresponding to a 3-site-per-unit-cell bipartite lattice with an $I\Ts$ symmetry that exchanges the two majority sublattices. Note this symmetry enforces the uniform pairing condition on the flat band.

Recall that in the main text, we define $\phi(k)=2\arg[f(k)]$, which we then decompose as $\phi(k)=\phi_0(k)+\alpha\tilde\phi(k)$, where $\phi_0(k)=-\phi_0(-k)$ (respecting $I$ and $\Ts$ separately) and $\tilde\phi(k)=\tilde\phi(-k)$ (breaking $I$ and $\Ts$ individually but preserving their product). We also require $\int\tilde\phi(k)dk=0$ (possible by $\alpha$-dependent choice of sublattice location) to minimize the extended quantum metric.

In terms of the extended-space vector $\Phi(k)=(\partial_{k_i}\phi_0(k)\ \tilde\phi(k))^T$, the extended quantum metric is given by the outer product $g=\Phi\Phi^T$. Notably, the conventional quantum metric is $g_{ij}=(\partial_{k_i}\phi_0)(\partial_{k_j}\phi_0)$, and the mixed component of the extended quantum metric that gives rise to the Lifshitz invariant is $g_{i\alpha}=\tilde\phi\partial_{k_i}\phi_0$. The resulting helical wavevector is
\begin{equation}
    q_0=-2\bar \alpha\left[\int(\vec{\partial}\phi_0)\otimes(\vec{\partial}\phi_0)dk\right]^{-1}\left[\int\tilde\phi\vec{\partial}\phi_0dk\right].
\end{equation}

Now consider the lattice illustrated in \cref{fig:1dbipartite}a. Letting $a_i$, $b_i$, $c_i$ denote annihilation operators on the three sites in unit cell $i$, the Hamiltonian takes the form
\begin{equation}\label{eq:bipartham}
    H=a_i^\dagger[t(e^{i\theta}b_i+e^{-i\theta}c_i)+t_+(b_{i-1}+c_{i+1})+t_-(b_{i+1}+c_{i-1})]+h.c.
\end{equation}
where $t$, $t'$, $t''$ are real and the phase $\theta$ is set by the magnetic flux. For this Hamiltonian,
\begin{equation}
    f(k)=te^{i\theta}+t_+e^{ika}+t_-e^{-ika}.
\end{equation}
We impose $t>t_++t_-$, as we desire a large gap at $\theta=0$. Note that if $t_+=t_-$ then the system has a vertical mirror symmetry at $\theta=0$, and the zero-energy states are the $-1$ eigenstate of this symmetry, which results in trivial quantum geometry in the flat band -- i.e., no superfluid weight.

Writing $t'=t_++t_-$ and $t''=t_+-t_-$, the phase of the wavefunction is
\begin{equation}
    \phi(k)=2\arctan(\frac{t\sin(\theta)+t''\sin(k)}{t\cos(\theta)+t'\cos(k)})-\bar\phi
\end{equation}
where $\bar\phi$ is set so that $\int \phi(k)=0$ to minimize the extended quantum metric. Then, we evaluate the integrals of quantum metric components over $k$ numerically. The value of $q_0$ as a function of $t_{\pm}/t$ is shown in \cref{fig:1dbipartite}c.

Aside from the $C_{2y}\Ts$ symmetry when $t_+=t_-$ imposing a vanishing Lifshitz invariant, the most notable feature of the helical vector $q_0$ is the limiting value  $q_0a \rightarrow \pm 2\theta$ when one of $t_\pm\rightarrow 0$. In this limit the model approaches an asymmetric diamond chain, as illustrated in \cref{fig:diamondchain}, and there is zero net flux through every loop in the lattice. As such, there is a gauge choice where every hopping is real; this gauge derives from putting the phase from the flux on the vanishing term of $t_{\pm}$ instead of $t_0$. However, as written, the hopping terms are complex, indicating that the time-reversal symmetry is not simply complex conjugation in this choice of gauge.

\begin{figure}
    \centering
    \begin{tikzpicture}[scale=2.5]

\begin{scope}[ultra thick,scale=0.75]
    \begin{scope}[dashed]
        \draw (-1,0) -- (0,0.75);
        \draw (-1,-0.75) -- (1,0.75);
        \draw (0,-0.75) -- (2,0.75);
        \draw (1,-0.75) -- (2,0);
    \end{scope}
    \begin{scope}[->,shorten >=2mm]
        \draw (0,-0.75) -- (0,0);
        \draw (0,0) -- (0,0.75);
        \draw (-1,-0.75) -- (-1,0);
        \draw (-1,0) -- (-1,0.75);
        \draw (1,-0.75) -- (1,0);
        \draw (1,0) -- (1,0.75);
        \draw (2,-0.75) -- (2,0);
        \draw (2,0) -- (2,0.75);
    \end{scope}
    \begin{scope}[gray]
        \filldraw(-1,0) circle (0.1cm);
        \filldraw (0,0) circle (0.1cm);
        \filldraw(1,0) circle (0.1cm);
        \filldraw(2,0) circle (0.1cm);
    \end{scope}
    \begin{scope}[red]
        \filldraw(-1,0.75) circle (0.1cm);
        \filldraw (0,0.75) circle (0.1cm);
        \filldraw(1,0.75) circle (0.1cm);
        \filldraw(2,0.75) circle (0.1cm);
        \filldraw(-1,-0.75) circle (0.1cm);
        \filldraw (0,-0.75) circle (0.1cm);
        \filldraw(1,-0.75) circle (0.1cm);
        \filldraw(2,-0.75) circle (0.1cm);
    \end{scope}

    \draw[->] (2.6,0.5) -- (3.2,0.5);
    \draw[dashed]  (2.6,-0.5) -- (3.2,-0.5);
    \node at (3.5,0.5) {$t_0e^{i\theta}$};
    \node at (3.5,-0.5) {$t_-$};
\end{scope}
\end{tikzpicture}
    \caption{In the limit where one of $t_{\pm}\rightarrow 0$, the model approaches an asymmetric diamond chain. Here $t_+$ is set to zero, and $t_0$ has a phase per \cref{eq:bipartham}.}
    \label{fig:diamondchain}
\end{figure}

\begin{figure}
    \centering
    \begin{tikzpicture}[ultra thick,scale=0.5]
\filldraw[fill=gray] (-3,3) -- (3,3) -- (3,-3) -- (2,-3) -- (2,2) -- (-2,2) -- (-2,-3) -- (-3,-3) -- cycle;
\filldraw[draw=blue,fill=blue!50] (-2,-3) -- (-2,-2) -- (2,-2) -- (2,-3) -- cycle;

\node[blue] at (0,-3.5) {\small $q_0$};
\draw[blue,->] (-1,-4) -- (1,-4);

\draw[|-|] (-2,3.5) -- (2,3.5);
\node at (0,4) {\small $L$};

\node[green!40!black] at (0,0) {\small $\phi_B+q_0L$};
\draw[green!40!black,->,thin] (1.8,0.2) -- (1.8,1.8) -- (-1.8,1.8) -- (-1.8,-1.8) -- (1.8,-1.8) -- (1.8,-0.2);

\draw[->] (0,-2.5) -- (6.5,-0.5);

\begin{scope}[shift={(10,0)},scale=3]
    \begin{scope}
        \draw[dashed] (-1,-0.75) -- (1,0.75);
        \draw[dotted] (1,-0.75) -- (-1,0.75);
        \draw[dashed] (0,0.75) -- (-1,0)  (0,-0.75) -- (1,0);
        \draw[dotted] (0,-0.75) -- (-1,0)  (0,0.75) -- (1,0);
        \draw (0,-0.75) -- (0,0.75);
        \draw (-1,-0.75) -- (-1,0.75);
        \draw (1,-0.75) -- (1,0.75);
    \end{scope}
    \begin{scope}[gray]
        \fill (-1,0) circle (0.1cm);
        \filldraw (0,0) circle (0.1cm);
        \fill (1,0) circle (0.1cm);
    \end{scope}
    \begin{scope}[red]
        \fill (-1,0.75) circle (0.1cm);
        \filldraw (0,0.75) circle (0.1cm);
        \fill (1,0.75) circle (0.1cm);
        \fill (-1,-0.75) circle (0.1cm);
        \filldraw (0,-0.75) circle (0.1cm);
        \fill (1,-0.75) circle (0.1cm);
    \end{scope}
    \begin{scope}[blue]
        \draw[<-] (-0.1,0.3) arc[start angle=255, end angle=-75, x radius=0.3, y radius=0.1];
        \draw[<-] (-0.1,-0.5) arc[start angle=255, end angle=-75, x radius=0.3, y radius=0.1];
    \end{scope}
    \draw[black] (0,-0.6) -- (0,-0.2) (0,0.2) -- (0,0.6);
\end{scope}

\draw[->] (9,-3.1) -- (8,-5.8);
\node at (10.1,-4.6) {$t_+\rightarrow 0$};

\begin{scope}[shift={(7,-9)},scale=3]
    \begin{scope}
        \draw[dashed] (-1,-0.75) -- (1,0.75);
        \draw[dashed] (0,0.75) -- (-1,0)  (0,-0.75) -- (1,0);
        \draw (0,-0.75) -- (0,0.75);
        \draw (-1,-0.75) -- (-1,0.75);
        \draw (1,-0.75) -- (1,0.75);
    \end{scope}
    \begin{scope}[gray]
        \fill (-1,0) circle (0.1cm);
        \filldraw (0,0) circle (0.1cm);
        \fill (1,0) circle (0.1cm);
    \end{scope}
    \begin{scope}[red]
        \fill (-1,0.75) circle (0.1cm);
        \filldraw (0,0.75) circle (0.1cm);
        \fill (1,0.75) circle (0.1cm);
        \fill (-1,-0.75) circle (0.1cm);
        \filldraw (0,-0.75) circle (0.1cm);
        \fill (1,-0.75) circle (0.1cm);
    \end{scope}
    \begin{scope}[blue]
        \fill[opacity=0.5] (-0.93,-0.1) -- (-0.93,-0.6) -- (-0.6,-0.35) -- cycle;
        \fill[opacity=0.5] (-0.07,0.1) -- (-0.07,0.6) -- (-0.4,0.35) -- cycle;
        \fill[opacity=0.5] (-0.93,0.1) -- (-0.93,0.6) -- (-0.6,0.35) -- cycle;
        \fill[opacity=0.5] (-0.07,-0.1) -- (-0.07,-0.6) -- (-0.4,-0.35) -- cycle;
        \node[rotate=35] at (0.5,0) {\tiny Zero net flux};
        \node at (-0.18,0.35) {$\bm{+}$};
        \node at (-0.18,-0.35) {$\bm{+}$};
        \node at (-0.82,-0.35) {$\bm{-}$};
        \node at (-0.82,0.35) {$\bm{-}$};
    \end{scope}
    \begin{scope}[green!40!black]
        \draw[thick,->] (-0.9,0.2) -- (-1.5,0.05);
        \node at (-2.15,0.15) {\small Flux inside};
        \node at (-2.15,-0.1) {\small SQUID};
    \end{scope}
\end{scope}

\end{tikzpicture}
    \caption{(Top left) RF SQUIDs made with one side a time-reversal-symmetric superconductor (gray) and the other a helical superconductor (blue) can measure the helical wavevector: the total phase around the green loop is the sum of the phase from the helical wavevector $q_0L$ and the magnetic flux through the loop $\phi_B$. (Top right) In the system of interest, the helical wavevector is the bipartite chain in \cref{fig:1dbipartite}. The magnetic flux can concretely be understood as field loops around the intracell hoppings, as indicated in blue. (Bottom) In the two limiting cases of this chain with vanishing diagonal hoppings, even though no magnetic flux passes through any hopping loop, some of the flux ends up inside the SQUID loop. This flux can be included in the tight binding hoppings of the 1d bipartite lattice, giving the helical wavevector computed in \cref{fig:1dbipartite}, or can be modeled as a net mesoscopic field contributing to the flux $\phi_B$ through the SQUID. Both approaches give the same net phase shift around the loop.}
    \label{fig:SQUID}
\end{figure}

To understand the diamond chain limit, consider connecting the helical superconductor in an RF SQUID, as illustrated in \cref{fig:SQUID}. The phase shift around the SQUID loop is the sum of the phase shift from magnetic flux through the loop and the helical wavevector times the length of the helical superconductor. We now show that continuously moving the magnetic field lines (i.e., without crossing field lines over hoppings) to get a time-reversal-symmetric chain necessarily results in a net flux through the SQUID that produces the same phase shift.

Consider modeling the magnetic flux within the bipartite lattice as magnetic fields circling around the $t_0$ hoppings and set one of the diagonal hoppings $t_{\pm}$ to zero, as illustrated in \cref{fig:SQUID}. The magnetic fluxes within the hopping loop cancel, so this magnetic field can be continuously reduced to zero without changing the flux. What remains is a magnetic field that rotates around the 1D chain as a whole - a positive flux below the chain and a negative flux above the chain, after setting $t_+$ to zero. Continuously pushing these fields away from the material allows these fields to be shifted out of our microscopics and into our mesoscopics: the flux is no longer viewed as a part of the tight-binding model, and instead viewed as magnetic flux through the SQUID.

Since no fluxes changed via this continuous process, the two methods should give the same result. The phase shift of the magnetic flux is the number of unit cells, $L/a$, times the flux in the upper triangle of \cref{fig:SQUID}, which is $2\theta$. This gives
\begin{equation}
    q_0L=\phi_B=(L/a)(2\theta)\quad\rightarrow\quad q_0a=2\theta
\end{equation}
in the $t_{\pm}\rightarrow 0$ limit, as found numerically in \cref{fig:1dbipartite}.

\section{Mixed quantum geometry describing deviations from quantum geometric nesting for charge, spin, and pair density waves}\label{Apx:DWstates}

The derivation of Ginzburg-Landau theory for charge and pair density waves is very similar to that of superconductors \cite{CompareCDWandSC}; in particular, aside from the coupled $k$ points and the time-reversal operator \mTs, the quadratic coefficient of free energy is virtually equivalent to \cref{eq:FreeEnResult}. For a flat-band charge density wave of wavevector $Q$, where $\Delta(q)=\langle c^\dagger_{k+Q+q}c_k\rangle$,
\begin{equation}
    \Omega^{(2)}=\sum_q|\Delta(q)|^2\left\{\frac{N}{U}-\frac{1}{8T}\sum_{k}\Tr[P_k V P_{k+Q+q}V^\dagger]\right\}.
\end{equation}
Here $V$ is a matrix representing the form of the charge (or spin, etc.) density wave. Similarly, for flat-band pair density waves, where $\Delta(q)=\langle c_{k+Q+q}c_{-k}\rangle$,
\begin{equation}
    \Omega^{(2)}=\sum_q|\Delta_0(q)|^2\left\{\frac{N}{U}-\frac{1}{8T}\sum_{k}\Tr[P_{-k}\mathcal{V} P_{k+Q+q}\mathcal{V}^\dagger]\right\},
\end{equation}
where $\mathcal{V}$ is antilinear (and $\mathcal{V}^\dagger$ is the antilinear adjoint).  For simplicity, we hereforward consider only a well-isolated perfectly flat band with projector $P_k$, and focus on the charge density wave state.

If there exist matrices $V$ or $\mathcal{V}$ such that the traces are maximized everywhere, then the state is maximally susceptible to a charge density wave state, a condition known as quantum geometric nesting \cite{QGNest}. For our purposes, we additionally require that the matrix $V$ be unitary (on the orbitals where the flat band has nonzero weight); this is the analogue of the uniform pairing condition in this context. In this case, maximizing the trace is equivalent to requiring that
\begin{equation}\label{eq:QGNcond}
    P_kVP_{k+Q}=P_kV
\end{equation} for all $k$ (with $V$ independent of $k$).  For pair density waves, we instead replace $V$ with $\mathcal{V}$ and require antiunitarity of $\mathcal{V}$.

One can derive the stiffness of the CDW phase by a procedure analogous to the superfluid weight by expanding to quadratic order in $q$. If quantum geometric nesting is perfect, the quantum metric appears in exactly the same form as in superconductivity, because the nesting condition reduces the trace in the free energy to $\Tr[P_kP_{k+q}]$ exactly as time-reversal symmetry does in the superconducting case.

Now we suppose that we do not have perfect quantum geometric nesting, but instead perturb away from it. The question of interest is: how does the resulting $Q$ change? The shift in $Q$ from the perfectly-nested value is the analogue of the helical wavevector in superconductivity, and both can be computed from the linear-in-$q$ correction, the Lifshitz invariant. Assuming the quantum geometric nesting vector $Q$ is commensurate, this shift in wavevector quantifies a commensurate-to-incommensurate transition of the CDW state.

Now, we proceed by analogy with the superconducting state to compute this quantum geometric Lifshitz invariant. Recall that to apply the perturbation theoretic formalism in the superconducting case, it is important that the Hamiltonian decomposes as
\begin{equation}
    H=H_0+\alpha H_1,\qquad [H_0,\Ts]=\{H_1,\Ts\}=0.
\end{equation}
The analogous condition here replaces time reversal $\Ts$ with the combination of the nesting matrix $V$ and the translation in momentum space, i.e., for the same decomposition $H=H_0+\alpha H_1$,
\begin{equation}
    H_0(k)=VH_0(k+Q)V^\dagger,\quad H_1(k)=-VH_1(k+Q)V^\dagger.
\end{equation}

Unlike the case of time-reversal, not all Hamiltonians will naturally decompose into such an $H_0$ and $H_1$. E.g., for a triangular lattice with a $\sqrt{3}\times\sqrt{3}$ charge density wave state, where $V=\I$ and $Q=K$, the requirement would imply
\begin{equation}
    H_1(k)=(-1)^3V^3H_1(k+3Q)(V^\dagger)^3=-H_1(k).
\end{equation}
Hence no perturbations to this particular CDW state could be analyzed in this way.

In general, whether there are perturbations that can be analyzed in this way depends on the eigenvalues of the nesting operator $Ve^{iQr}$. In general, only eigenvalues of equal magnitude but opposite sign admit an anticommuting matrix. The above calculation indicates that the $\sqrt{3}$ CDW state only has the cube roots of negative unity as eigenvalues of $Ve^{iQr}$, hence no matrix anticommutes with it as required.

By contrast, for e.g. a $(\pi,\pi)$ CDW state on a square lattice, the only eigenvalues are $\pm 1$, hence any Hamiltonian is decomposable into commuting and anticommuting parts like the superconducting case. For most CDW or PDW states, the set of anticommuting perturbations will be a subspace of the set of all perturbations - i.e., some kinds of perturbations can be analyzed with this mixed quantum geometry approach and others cannot.

In the end, once such a perturbation is identified, the analysis is virtually identical to the superconducting case: the mixed quantum geometry describes the Lifshitz invariant, hence the linear shift in the preferred $Q$.

\section{Nonstandard order parameters}\label{Apx:OtherOrders}
In the rest of this text, we assume that the form of the interaction is entirely on-site, meaning that the resulting pairing $\Delta$ is also on-site. In this section we show how in the presence of interactions that allow intersite pairing (subject to a suitable analogue of the uniform pairing condition), mixed quantum geometry can shed some light on the amount of intersite pairing.

Suppose we have an intra-unit-cell density-density interaction which for simplicity is still only between opposite spins, $U_{\alpha\beta}n_{i\alpha\uparrow}n_{j\beta\downarrow}$ where $ij$ run over unit cells and $\alpha\beta$ run over sublattices and orbitals. Conventional on-site interactions would be of the form $U_{\alpha\beta}=\delta_{\alpha\beta}$; we will instead consider $U_{\alpha\beta}=U$ (i.e., the same strength between all orbitals in a unit cell).

This means that the full matrix structure of $\hat\Delta_{ij}$ must be considered; we now impose a new restriction that generalizes the uniform pairing condition: we assume $\hat\Delta$ is a multiple of a unitary matrix, $\hat\Delta=\Delta_0 V$. This is done without thought for physical motivation in any particular case; it is simply necessary for the mathematical approach we present here to work. 

Taking an inversion-symmetric exactly flat band for simplicity and focusing for now on the $q=0$ state (the ground state, as the Lifshitz invariant vanishes) the free energy in this case is
\begin{equation}
    \Omega^{(2)}=\Delta_0^2\{U^{-1}N-(8T)^{-1}\sum_k\Tr[P_{k,\alpha} VP_{k,-\alpha}V^\dagger]\}.
\end{equation}
The trace must now be optimized over all possible choices of $V$.

This procedure was already performed for special choices of $V$ in \cref{Apx:MinQuantMet} by taking the minimal quantum metric. The key insight to that derivation, which can also be applied here, is that for a fixed $q=q_0$, $V$ can be absorbed into the projectors. Here, we write $V=\exp(i\bar\alpha M/2)$ with $M$ some Hermitian matrix, whereupon in terms of the projectors $\tilde P(k,\alpha)=e^{i\alpha M}P_{k,\alpha}e^{-i\alpha M}$, the trace reduces to $\Tr[P_{k,\alpha}P_{k,-\alpha}]$ as before.

To optimize over these new parameters, it is convenient to introduce parameters $c_m$ such that $M=\sum_m c_m\lambda_m$, where $\lambda_m$ span appropriately-sized Hermitian matrices. Minimizing the original free energy over all forms of $\Delta$ now corresponds to minimizing over these choices of $c_m$.

These projectors are the flat-band states of a modified Hamiltonian
\begin{equation}
    \tilde H(k,\alpha,c)=e^{i\sum c_m\lambda_m}H(k,\alpha)e^{-ic_m\lambda_m}.
\end{equation}
We now note that the same procedure used to optimize over $q$ can here be used to optimize over $c_m$; one ends up with a purely parameter space quantum metric with indices that run over $\alpha$ and $m$. Note some care must be taken for spin triplet systems to maintain Pauli exclusion: $V$ should be restricted to antisymmetric matrices when spin space is included.

To work in a system with both a Lifshitz invariant and some freedom in the order parameter, the extended quantum metric must now include all three parts: the momentum $q$, the time-reversal-breaking $\alpha$, and the order parameter form $c$. To compute the helical wavevector, minimize $q$ and $c$ for fixed $\alpha$; to then compute the superfluid weight, minimize only $c$ at fixed $q$ and $\alpha$, then expand about the helical wavevector to quadratic order.

\end{document}